\documentclass[symmetry,article,accept,pdftex,moreauthors]{Definitions/mdpi}

\firstpage{1} 
\pubvolume{1}
\issuenum{1}
\articlenumber{0}
\pubyear{2024}
\copyrightyear{2024}
\externaleditor{Academic Editor: Andrea Lavagno}
\datereceived{19 December 2023} 
\daterevised{18 January 2024} 
\dateaccepted{22 January 2024} 
\datepublished{ } 
\hreflink{https://doi.org/} 

\usepackage[]{mdframed}
\usepackage{amsfonts, amsthm, tensor, empheq,subcaption,lipsum}
\usepackage[export]{adjustbox}

\newcommand{\pdrv}[2]{\frac{\partial #1}{\partial #2}}
\newcommand{\tpdrv}[2]{\tfrac{\partial #1}{\partial #2}}

\newcommand{\eps}{\varepsilon}

\newcommand{\ovl}[1]{{\overline{#1}}}

\newcommand{\op}{\operatorname}

\usepackage[labelformat=simple]{subcaption}

\DeclareCaptionLabelFormat{subcaptionlabel}{\normalfont(\textbf{#2}\normalfont)}
\usepackage{dsfont}

\Title{Revisiting de Broglie's Double-Solution Pilot-Wave Theory with a Lorentz-Covariant Lagrangian Framework}

\TitleCitation{Revisiting de Broglie's Double-Solution Pilot-Wave Theory with a Lorentz-Covariant Lagrangian Framework}

\Author{David Darrow\orcidA{} and  John W. M. Bush *\orcidB{}}

\AuthorNames{David Darrow and John W. M. Bush}

\AuthorCitation{Darrow, D.; Bush, J.W.M.}

\address[1]{%
Department of Mathematics, Massachusetts Institute of Technology, Cambridge, MA 02139, USA; ddarrow@mit.edu
}

\corres{\hangafter=1 \hangindent=1.05em \hspace{-0.82em} Correspondence: bush@math.mit.edu}

\abstract{The relation between de Broglie's double-solution approach to quantum dynamics and the hydrodynamic pilot-wave system has motivated a number of recent revisitations and extensions of de Broglie's theory. Building upon these recent developments, we here introduce a rich family of pilot-wave systems, with a view to {reformulating and studying de Broglie’s double-solution program in the modern language of classical field theory}. Notably, the entire family is local and Lorentz-invariant, follows from a variational principle, and exhibits time-invariant, two-way coupling between particle and pilot-wave field. We first introduce a variational framework for generic pilot-wave systems, including a derivation of particle-wave exchange of Noether currents. We then focus on a particular limit of our system, in which the particle is propelled by the local gradient of its pilot wave. In this case, we see that the Compton-scale oscillations proposed by de Broglie emerge naturally in the form of particle vibrations, and that the vibration modes dynamically adjust to match the Compton frequency in the rest frame of the particle. The underlying field dynamically changes its radiation patterns in order to satisfy the de Broglie relation $p=\hbar k$ at the particle's position, even as the particle momentum $p$ changes. The wave form and frequency thus evolve so as to conform to de Broglie's \emph{harmony of phases}, even for unsteady particle motion. We show that the particle is always dressed with a Compton-scale Yukawa wavepacket, independent of its trajectory, and that the associated energy imparts a constant increase to the particle's inertial mass. Finally, we see that the particle's wave-induced Compton-scale oscillation gives rise to a classical version of the Heisenberg uncertainty principle.}

\keyword{Klein--Gordon equation; hydrodynamic quantum analogues; pilot-wave theory; \linebreak  Zitterbewegung; Heisenberg uncertainty principle; harmony of phases; Lagrangian mechanics} 

\begin{document}


\section{Introduction}

In 1923, prior to the advent of our modern approach to quantum mechanics, de Broglie proposed 
a physical picture of quantum dynamics~\cite{deBroglie1923,deBroglie1930}. 
This picture was based on a fundamental symmetry argument. 
Specifically, at that time, light was understood to exhibit both particle and wave aspects; he proposed that matter, too, must share this dual nature. 
According to de Broglie's so-called double-solution theory~\cite{deBroglie1956,deBroglie1970,deBroglie1987}, microscopic quantum particles have an internal vibration,  
which acts as a source of waves that serve to guide or `pilot' the particle. The wave-particle coupling was constrained by de Broglie's \emph{harmony of phases}, a principle he referred to as a ``grand loi de la nature'': the particle and wave vibration are always in synchrony, locked in phase.
The resulting pilot-wave dynamics 
was then posited, but never proven, to give rise to an emergent statistical behavior consistent with the standard predictions of quantum mechanics. 

De Broglie's double-solution theory was not without its successes. On the basis of his physical picture, he predicted electron diffraction, the 
validation of which by Davisson and Germer~\cite{Davisson1927} led to de Broglie's
Nobel Prize in 1929, awarded him ``for his discovery of the wave nature of electrons''. 
His theory led to a number of cornerstones of modern quantum mechanics. For example, the frequency of particle vibration 
was deduced by equating the particle's rest mass energy $E = mc^2$ with wave energy $\hbar \omega$,
yielding the so-called Einstein-de Broglie relation and the Compton frequency 
$\omega_c = mc^2/\hbar$. His physical picture also led naturally to the deduction of the de Broglie relation, $p = \hbar k$, as follows from equating the particle velocity with the group velocity of its Klein--Gordon pilot-wave field. 
Despite its early successes, the theory was never satisfactorily completed; consequently, it never took center stage in the development of quantum theory.


It is important to clarify the distinction between de Broglie's double-solution theory and the relatively well-known \emph{Bohmian mechanics}, or \emph{de Broglie--Bohm pilot-wave theory}~\cite{Bohm1952a,Holland1993,DurrBook2}. According to the latter, quantum particles are guided by the standard wave function; however, the particles are not seen as sources of that wave, whose form is unaltered by the particle position. Recent advances in Bohmian mechanics include the Lagrangian theories of Sutherland \cite{Sutherland_2015} and Holland \cite{Holland2020}. These Lagrangian approaches bear some similarity to the present work; indeed, both theories fall under the purview of the variational results to be developed in Section~\ref{sec:model}. However, our focus is on a classical pilot-wave dynamics of the form envisaged in de Broglie's double-solution theory and engendered in the walking-droplet system, where the particle responds exclusively to a wave of its own making. 


De Broglie's double-solution theory had two principle shortcomings. First, the physical nature of the wave
was never specified.
A number of possibilities have since been proposed and explored. 
The most highly developed pilot-wave theory of the form proposed by de Broglie may be found in stochastic electrodynamics~\cite{Milonni1994,Boyer2018}, in the work of de la Pe\~{n}a and Cetto~\cite{Pena1996,Pena2015}, who seek 
matter waves in the electromagnetic quantum vacuum.  The possibility of matter waves 
being of gravitational origin has also been explored, in which case they would represent undulations 
in the fabric of spacetime~\cite{Feoli1998, Derrico2023}. The second shortcoming of de Broglie's double-solution theory is that the manner in which the proposed 
pilot-wave dynamics could give rise to statistics of the form arising in quantum systems was never made clear. However, evidence of quantum-like statistics emerging from classical pilot-wave dynamics has come from recent advances in fluid mechanics.

In 2005, Couder and Fort discovered a hydrodynamic pilot-wave system in which a millimetric droplet
self-propels over the surface of a vibrating bath, piloted through a resonant interaction with its own
wave field~\cite{Couder2005a}. The resulting `walker' comprises both the droplet and its quasi-monochromatic wave
field, and so represents a macroscopic realization of wave-particle duality~\cite{Couder2005a,Protiere2006}.
Remarkably, this system has captured a number of features of quantum systems previously
thought to be unique to the microscopic realm~\cite{Bush2015a}. Consequently, it has launched the field of hydrodynamic quantum
analogues~\cite{BushOza2020}, the goal of which is to revisit and redefine the boundary between classical and quantum behaviours. 
The map between de Broglie's double 
solution theory and the walking-droplet system is direct: the bouncing of the droplet at the Faraday frequency plays the
role of de Broglie's particle vibration at the Compton frequency, and the Faraday wavelength the role of the de Broglie wavelength~\cite{Bush2015a}.

The hydrodynamic pilot-wave system has made clear the richness and complexity of classical pilot-wave
dynamics, as could not have been anticipated by physicists a century ago. In particular, it has shown 
how classical pilot-wave dynamics of the form proposed by de Broglie may give rise to emergent 
statistics of the form described by quantum mechanics. Salient examples include the 
hydrodynamic analogues of single-particle diffraction and interference~\cite{Couder2006,Pucci2018,Ellegaard2020}, quantised orbits 
~\cite{Couder2005a,Fort2010,Oza2014a,Perrard2014,Labousse2016,Durey2017}, the quantum 
corral~\cite{Harris2013a}, statistical projection effects~\cite{Saenz2018}, Friedel oscillations~\cite{Saenz2019a}, 
Anderson localisation~\cite{Saenz2023}, and surreal trajectories~\cite{frumkin_real_2022}. 
A remarkable feature of the hydrodynamic system is that the three fundamental timescales 
in the problem, that of particle vibration (0.01s), particle translation (1s), and statistical convergence 
(1h), may all be readily resolved in the laboratory~\cite{Bush2015a,BushOza2020}. 

The successes of pilot-wave hydrodynamics in capturing quantum phenomena have 
motivated and informed the exploration of a broader class of classical pilot-wave \linebreak  systems~\cite{Grossing2015,Drezet2020,Borghesi2017,Durey2021}. 
Moreover, they have motivated a revisitation of de Broglie's double-solution theory~\cite{Colin2017,Hatifi2018,Drezet2023,Drezet2023b}. 
{In their recent research program, the so-called \emph{hydrodynamic quantum field theory} (HQFT),} Dagan and Bush~\cite{Dagan2020}, Durey and Bush~\cite{Durey2020b}, and Dagan~\cite{Dagan2022} considered particles with an intrinsic vibration 
generating, then moving in response to, a Klein--Gordon field. This pilot-wave field was periodically forced at twice the Compton frequency in a finite region adjoining the particle. 
For the sake of simplicity, particle inertia was neglected: the particle velocity was taken to be proportional to the local wave gradient and the trajectory equation was first-order in time.
The coupling constant between particle velocity and wave gradient
was determinant in the resulting dynamics: above a critical value, the particle
self-propelled~\cite{Dagan2020}. The pilot wave form was marked by radial waves with the 
Compton wavelength radiating energy outwards from the particle path, plus a plane wave with the de Broglie wavelength moving in synchrony with the particle.
The free particle was found to follow an irregular, quasi-random walk, with a mean momentum $\langle p^2\rangle = \hbar^2k^2$ prescribed by the coupling constant. Moreover, the free particle 
motion was marked by a smaller-scale, erratic component, reminiscent
of the Zitterbewegung posited for the Dirac electron~\cite{1990FoPh...20.1213H}.

The question naturally arises: how might Louis de Broglie's double-solution theory have evolved 
if he had the computational facilities available to us today, and the physical picture furnished
by pilot-wave hydrodynamics? We take another step toward answering this question by 
presenting a new class of dynamical systems that represent a modern version of de Broglie's 
double-solution theory, offering a Lorentz-covariant, Lagrangian formulation of classical pilot-wave dynamics.
{We emphasize that our work is not an attempt to refute or modify modern quantum mechanics. 
Rather, we aim to reformulate and study de Broglie's double-solution program in the modern language 
of classical field theory.} 



In Section~\ref{sec:model}, we present the mathematical model under consideration in its most general form. Specifically, we develop a relativistic, variational theory for a point-particle and a scalar (Klein--Gordon) field, described in terms of Euler--Lagrange equations. 
In Section~\ref{sec:amplitude}, we introduce a specific, \emph{amplitude modulated} (AM) limit of our system, wherein the particle moves in response to the local gradient of the pilot wave. After developing a version of Noether's theorem (and expressions for the particle-wave exchange of Noether currents) in the general case, we study the AM system analytically in Sections~\ref{sec:AMconservation} and \ref{sec:steadystate}, characterizing the system's energetics
and showing that a trajectory-invariant, Compton-scale wavepacket is maintained about the particle at all times. In Sections~\ref{sec:oscillation} and \ref{sec:radiation}, we present our numerical study of the system, which reveals a new, \emph{dynamical} version of de Broglie's harmony of phases. 
In Section~\ref{sec:virtual}, we 
characterize the energy partitioning between the particle and its immediate surroundings,
as well as the manner in which the pilot wave alters the particle's inertial response.
Finally, we demonstrate in Section~\ref{sec:uncertainty} that, in wall-bounded geometries, the particle oscillation and effective mass combine to give rise to a classical version of Heisenberg's uncertainty principle for position and momentum measurements made in any direction.

\section{Mathematical Model}\label{sec:model}
In the following section, we introduce our proposed pilot-wave model in its most general form, beginning with a Lagrangian framework. Except where stated otherwise, we apply natural units $\hbar = c = 1$. For all tensor indices, we use Greek indices ($\mu$, $\nu$, etc.) to run over spacetime indices $(0,1,2,3)$, and Roman indices ($j$, $k$, etc.) to run over space indices $(1,2,3)$. We apply the ``mostly minuses'' convention for the metric tensor: $\eta^{\mu\nu}\sim(+,-,-,-)$, so the d'Alembert operator is $\partial^\mu\partial_\mu = \partial_{t}^2 - \nabla^2$. Finally, we use the vector notation $\vec{q}$ for three-vectors; if one exists, the corresponding four-vector is simply denoted without the overline, as $q$.

Our system is made up of a complex, free Klein--Gordon field $\phi$ of mass density $\tilde{m}$ and a relativistic particle of mass $m$ at the point $q_p\in\mathbb{R}^3$, coupled to $\phi$ through a real-valued coupling function $\sigma(\phi(q_p),\phi^*(q_p))=:\sigma(q_p)$. The total action $\mathcal{S}$ may be expressed as the sum of the actions of a free field and of a particle of inertial mass $\sigma$:

\begin{equation}\label{eq:action}
	\begin{gathered}
	\mathcal{S} = \mathcal{S}_\text{field} + \mathcal{S}_\text{part.},\\
	\mathcal{S}_\text{field}= \int d^4q\;m^2\left(\partial^\mu\phi^*\partial_\mu\phi - \tilde{m}^2|\phi|^2\right),\\
	\mathcal{S}_\text{part.} = -\int_0^{t'} dt\;\gamma^{-1}m\sigma(\phi,\phi^*),
	\end{gathered}
\end{equation}
where $\gamma = (1 - (\frac{\vec{u}}{c})^2)^{-1/2}$ is the Lorentz factor of the particle, and $\vec{u}$ its velocity. The normalization $m^2$ of $\mathcal{S}_\text{field}$ is chosen to ensure that $\phi$ is dimensionless.

Stepping from the action (\ref{eq:action}) to appropriate dynamical equations requires a form of the \emph{Euler--Lagrange equations} for pilot-wave systems, which we establish in Lemma \ref{lem:eullag} below. While similar derivations have likely been performed in the context of classical electromagnetism, they are not known to the authors. The proof of the following result is given in Appendix~\ref{sec:variationproofs_A}.

\begin{Lemma}[Euler--Lagrange equations for a pilot-wave]\label{lem:eullag}
	Imagine that our action takes the general form (\ref{eq:action}) with
	\begin{align*}
		\mathcal{S}_\text{field} &= \int d^4q \;\mathcal{L}_\text{field}(\phi,\partial_\mu\phi, q),\\
		\mathcal{S}_\text{part.} &= \int dt\;\mathcal{L}_\text{part.}(q_p,u,\tau,\phi,\partial_\mu\phi),
	\end{align*}
	where $u = \dot{q}_p$ {and $\tau$ is the proper time along the particle's trajectory}. This action is extremized when $\phi$ and $q_p$ satisfy the following equations of motion:
 \begin{adjustwidth}{-\extralength}{0cm}
	\begin{equation}\label{eq:lem1eqs}
        \begin{gathered}
		d_t\delta_{\vec{u}}\mathcal{L}_\text{part.} = \delta_{\vec{q}_p}\mathcal{L}_\text{part.},\\
		(\partial_\mu\delta_{\phi_\mu} - \delta_\phi)\mathcal{L}_\text{field} = -\delta^3(\vec{q}-\vec{q}_p)(\partial_\mu\delta_{\phi_\mu} - \delta_\phi)\mathcal{L}_\text{part.}+ \nabla\delta^3(\vec{q}-\vec{q}_p)\cdot (\vec{u}\delta_{\phi_0} -\delta_{\nabla\phi})\mathcal{L}_\text{part.},
        \end{gathered}
	\end{equation}
 \end{adjustwidth}
	where we interpret derivatives of $\delta^3$ {in a distributional sense} (i.e., given a distribution $\varphi:C^\infty_0\to\mathbb{R}$, the derivative $\partial\varphi$ is defined by $(\partial\varphi)(f) = -\varphi(\partial f)$), and where $\delta_{\nabla\phi}$ represents the three-vector of variational derivatives $\delta_{\partial_k\phi}$.
	Note that when $\delta_{\nabla\phi}\mathcal{L}_\text{part.} = \vec{u}\delta_{\phi_0}\mathcal{L}_\text{part.}$, the term proportional to $\nabla\delta^3$ vanishes.
\end{Lemma}
\begin{Remark}
	At first glance, it appears that the term $\partial_\mu\delta_{\phi_\mu}\mathcal{L}_\text{part.}$ is not well-defined; namely, as it is always multiplied by the delta function, we can interpret $\mathcal{L}_\text{part.}$ equally well as either a function of space coordinates $\vec{q}$ or of particle position $\vec{q}_p = \vec{q}_p(t)$. However, we see that this distinction is inconsequential: by grouping it with the term $\nabla\delta^3(\vec{q}-\vec{q}_p)\cdot(\vec{u}\delta_{\phi_0}-\delta_{\nabla\phi})\mathcal{L}_\text{part.}$ and integrating against a test function $\psi$, we see that
 \begin{adjustwidth}{-\extralength}{0cm}
    \vspace{-\baselineskip}
	\begin{align*}
		\int d^3\vec{q}\;\psi(\vec{q}&)\big(\delta^3(\vec{q}-\vec{q}_p)\partial_\mu\delta_{\phi_\mu}\mathcal{L}_\text{part.} + \nabla\delta^3(\vec{q}-\vec{q}_p)\cdot(\delta_{\nabla\phi}-\vec{u}\delta_{\phi_0})\mathcal{L}_\text{part.}\big)=-\left(\partial_\mu\psi(\vec{q})\right) \delta_{\phi_\mu}\mathcal{L}_\text{part.},
	\end{align*}
 \end{adjustwidth}
	which does not depend on spacetime derivatives of $\mathcal{L}_\text{part.}$.
\end{Remark}

Returning to the action (\ref{eq:action}) and extremizing against $\phi^*$ and $\vec{u}$, as in Lemma \ref{lem:eullag}, gives rise to the following Lorentz-covariant equations governing the evolution of the wave and particle, respectively:
\vspace{-6pt}
\begin{empheq}[box=\fbox]{equation}\label{eq:oureqs}
	\begin{gathered}
		(\partial_\mu\partial^\mu + \tilde{m}^2)\phi = m^{-1}\gamma^{-1}\delta^3(\vec{q}-\vec{q}_p)\pdrv{\sigma}{\phi^*}\\
		d_t\left(m\sigma\gamma\vec{u}\right) = \gamma^{-1}m\nabla\sigma(q_p)
	\end{gathered}
\end{empheq}

There are a few key differences between our formalism and the {\emph{HQFT} program} of Dagan and Bush \cite{Dagan2020}. First, and most notably, HQFT's guiding equation is only first order in the particle position: particle inertia was neglected, precluding its reduction
to classical mechanics in the limit of $\hbar \rightarrow 0$. Our Lagrangian formalism forces the guiding equation to be \emph{second} order, giving rise to particle inertia and the appropriate classical limit. Second, the system (\ref{eq:oureqs}) is exactly Lorentz-covariant, where HQFT depends on the current reference frame. As we shall see, this difference has a profound influence on the form of the \linebreak  pilot wave.

Finally, in HQFT,
a forcing of the form $\cos(2\omega_c t)$ is imposed over a finite region (of the Compton scale) 
by the particle on the field.
In HQFT, this coupling gives rise to a particle vibration reminiscent of the \emph{Zitterbewegung} of the Dirac theory \cite{1990FoPh...20.1213H}.
%
In Section~\ref{sec:oscillation}, we will demonstrate that a similar Zitterbewegung emerges naturally (without being imposed) from the equations (\ref{eq:oureqs}); moreover, the oscillation frequency adjusts dynamically to align with $\omega_c$ in the \emph{rest frame} of the particle.

Another nuance of (\ref{eq:oureqs}) is that the coupling function $\sigma$ contributes directly to the particle's inertial mass. A necessary artifact of Lorentz covariance, this mass-coupling ensures that the particle remains on-shell (satisfying $E^2 = m^2+p^2$) at all times; moreover, it enables the particle and field to exchange energy at the point $\vec{q}_p$. 


\subsection{Defining the Field at the Particle Position}\label{sec:continuous}

We now address the outstanding problem with the system (\ref{eq:oureqs}); specifically, while the field $\phi$ is generally singular at the point $\vec{q}_p$, our action depends on the value $\phi(\vec{q}_p)$.
In the particular system considered in Section~\ref{sec:amplitude}, for instance, the field locally takes the form of a Yukawa potential about the particle: $\phi\propto e^{-mr}/r$. 
To place our theory on firm mathematical footing and sidestep this singularity, we project the field onto its \emph{continuous component} at the \linebreak  point $\vec{q}_p$:
\begin{Definition}
	Given a function $\phi:\mathbb{R}^3\times\mathbb{R}\to\mathbb{C}$, define the \emph{continuous component} $\ovl{\phi}$ of $\phi$ at a point $(t,q)$ as
	\begin{equation}\label{eq:contcomponent}
	    \ovl{\phi}(t,q) = \lim_{r\to 0}\frac{1}{4\pi}\int_{S^2}d\xi\; \partial_r(r\phi(t,\vec{q} + r\vec{\xi})),
	\end{equation}
	where $\xi$ parametrizes $S^2\subset\mathbb{R}^3$. 
	When $\phi$ is continuous, we have $\phi = \ovl{\phi}$; however, if 
	\begin{equation}\label{eq:singularity}
		\phi(t,\vec{q}) = \phi_1(t,\vec{q}) + \frac{a}{\sqrt{(\vec{q}-\vec{q}_p)^T A (\vec{q}-\vec{q}_p)}}
	\end{equation}
	near a continuous trajectory $q_p=q_p(t)$, for a positive-definite, symmetric matrix $A$ and a continuous $\phi_1$, then we instead find $\ovl{\phi}(q_p) = \phi_1(q_p)$. This decomposition demonstrates that the construction (\ref{eq:contcomponent}) is Lorentz-invariant when restricted to functions of this form; indeed, any Lorentz transformation takes the singular term of (\ref{eq:singularity}) to another of the same form.
\end{Definition}

Notably, the Yukawa potential is of the form (\ref{eq:singularity}) in all reference frames, so the continuous component is Lorentz-invariant in the setting of Section~\ref{sec:amplitude}. With this construction in hand, we clarify that the coupling $\sigma$ in the action (\ref{eq:action}) must be of the form
\[\sigma(\phi,\phi^*) = \sigma(\ovl{\phi},\ovl{\phi}^*).\]
\textls[-10]{Because the projection $\phi\mapsto\ovl{\phi}$ is linear and preserves continuous perturbations, this redefinition does not affect the variational result (\ref{eq:oureqs}). It also allows the self-consistent definition}
\[\nabla\sigma(\phi,\phi^*) = \nabla\sigma(\phi_1,\phi_1^*),\]
where $\phi_1$ is defined in a neighborhood of $q_p$ by (\ref{eq:singularity}). Equivalently, for functions of that form, this expression is the mean gradient over a ball of radius $r\to 0$. 


We note that the construction (\ref{eq:contcomponent}) does \emph{not} solve the long-standing problem of defining the self-energy of a classical point-particle, e.g., as discussed by Hammond~\cite{Hammond2010}. In particular, it does not give a consistent, finite integral of energy over the full field. Nevertheless, it does provide a rigorous mathematical footing for the derivation of the Euler--Lagrange Equation (\ref{eq:oureqs}).

\subsection{Noether's Theorem for the Pilot-Wave System}\label{sec:noether}

With the application to future pilot-wave theories in mind, we leverage Lemma \ref{lem:eullag} to deduce a version of Noether's theorem for a general pilot-wave setting, in Lemma \ref{lem:noether} below. Through Corollary \ref{cor:exchange}, we also quantify the exchange of a general Noether current between the particle and wave. 
Note that the following results apply equally well to any Lagrangian pilot-wave theory, including those proposed by Sutherland \cite{Sutherland_2015} and Holland \cite{Holland2020}.

First, we introduce the following non-conservative version of the Euler--Lagrange equations (\ref{eq:lem1eqs}):
\begin{equation}\label{eq:noncons}
    d_t\delta_{\vec{u}}\mathcal{L}_\text{part.} - \delta_{\vec{q}}\mathcal{L}_\text{part.} = \vec{F},
\end{equation}
where $\vec{F} = \vec{F}(q_p,u,t,\phi,\partial_\mu\phi)$ is a given vector field. As in traditional Lagrangian mechanics, we can think of $\vec{F}$ as a non-conservative force applied to the particle directly.

\begin{Lemma}[Noether's theorem for a pilot-wave]\label{lem:noether}
	Let $\mathcal{L}_\text{field}$ and $\mathcal{L}_\text{part.}$ be as in Lemma \ref{lem:eullag}. Suppose we have a transformation
	\[(q_t)^k\mapsto (q_t)^k + \eps Q^k,\qquad \phi\mapsto \phi + \eps\Psi,\]
	written to the first order in $\eps$, and suppose that the Lagrangian densities transform as follows:
	\begin{align*}
		\mathcal{L}_\text{field} \mapsto \mathcal{L}'_\text{field}&:=\mathcal{L}_\text{field}+ \eps\partial_\mu\Lambda_\text{field}^\mu,\\
		\mathcal{L}_\text{part.} \mapsto \mathcal{L}'_\text{part.}&:=\mathcal{L}_\text{part.} + \eps d_t\Lambda_\text{part.},
	\end{align*}
	where 
 \begin{align*}\Lambda_\text{field}^\mu&=\Lambda_\text{field}^\mu(\psi,\partial_\mu\psi,q),\\
		\qquad \Lambda_\text{part.}&=\Lambda_\text{part.}(q_p,u,t,\phi,\partial_\mu\phi).
	\end{align*}
	In this setting, the quantity
	\begin{align*}
		j^\mu:=\Lambda^\mu_\text{field} - (\delta_{\phi_\mu}\mathcal{L}_\text{total})\Psi + u^\mu\delta^3(\vec{q}-\vec{q}_p)(\Lambda_{\text{part.}} - (\delta_{u^k}\mathcal{L}_\text{part.})Q^k)
	\end{align*}
	is conserved in the Euler--Lagrange evolution of Lemma \ref{lem:eullag}. That is,
	\[\partial_\mu j^\mu = 0,\]
	with derivatives interpreted {in a distributional sense (see the note in Lemma \ref{lem:eullag})}. If the particle evolution is instead replaced with the non-conservative version (\ref{eq:noncons}) of Lagrange's equations,
	we find that
	\[\partial_\mu j^\mu = -\delta^3(\vec{q}-\vec{q}_p)F_kQ^k.\]
\end{Lemma}
\begin{Remark}
	We note that this framework includes passive transformations of the coordinate system, simply by writing them with respect to the original coordinates. In general, a spacetime transformation $q^\mu\mapsto q^\mu+ \eps Q^\mu(q)$ is given by
	\[(q_t)^k\mapsto (q_t)^k + \eps (Q^k - u^kQ^0),\qquad \phi \mapsto \phi - \eps Q^\mu\partial_\mu\phi.\]
\end{Remark}

Following directly from the proof of this lemma (given in Appendix~\ref{sec:variationproofs_A}), we can quantify the exchange of any Noether current from particle to field and vice versa:

\begin{Corollary}\label{cor:exchange}
	In the setting of Lemma \ref{lem:noether}, define $j^\mu_\text{sing}$ and $j^\mu_\text{field}$ as
	\[j^\mu_\text{part.} = u^\mu\delta^3(\vec{q}-\vec{q}_p)(\Lambda_\text{part.}-(\delta_{u^k}\mathcal{L}_\text{part.})Q^k - (\delta_{\phi_\mu}\mathcal{L}_\text{part.})\Psi),\]
	\[j^\mu_\text{field} = \Lambda^\mu_\text{field} - (\delta_{\phi_\mu}\mathcal{L}_\text{field})\Psi.\]
	Then $j^\mu = j^\mu_\text{part.} + j^\mu_\text{field}$, and we have the following balance laws:
    \begin{adjustwidth}{-\extralength}{0cm}
  \vspace{-\baselineskip}
	\begin{gather*}
	    \partial_\mu j^\mu_\text{part.} = \delta^3(\vec{q}-\vec{q}_p)((\delta_\phi\mathcal{L}_\text{part.}-\partial_\mu\delta_{\phi_\mu}\mathcal{L}_\text{part.})\Psi -F_kQ^k) + \Psi\nabla\delta^3(\vec{q}-\vec{q}_p)\cdot (\vec{u}\delta_{\phi_0} -\delta_{\nabla\phi})(\mathcal{L}_\text{part.}),\\
     \partial_\mu j^\mu_\text{field} = -\delta^3(\vec{q}-\vec{q}_p)(\delta_\phi\mathcal{L}_\text{part.}-\partial_\mu\delta_{\phi_\mu}\mathcal{L}_\text{part.})\Psi - \Psi\nabla\delta^3(\vec{q}-\vec{q}_p)\cdot (\vec{u}\delta_{\phi_0} -\delta_{\nabla\phi})(\mathcal{L}_\text{part.}).
	\end{gather*}
    \end{adjustwidth}
\end{Corollary}

We will return to these results in Section~\ref{sec:AMconservation}, where we derive a conservation of stress-energy and of angular momentum for the system of interest.

\section{Amplitude-Modulated Dynamics}\label{sec:amplitude}
With these general results in hand, we restrict attention now to the \emph{amplitude-modulated} limit of our system (\ref{eq:oureqs}), defined by 
\[\sigma = 1 + b^2/4\pi + b\op{Re}\phi, \qquad\tilde{m} = m,\]
for a chosen coupling constant $b$. The $b^2/4\pi$ term serves to offset other, constant terms from the continuous-component construction (\ref{eq:contcomponent}), as is further discussed in Appendix~\ref{sec:particle_equation_A}.

This choice of $\sigma$ decouples the imaginary component of $\phi$ from the particle, so we can simply identify $\phi = \op{Re}\phi$. This field thus satisfies the modified equation
\vspace{-3pt}
\begin{empheq}[box=\fbox]{equation}\label{eq:equations_am1}
	(\partial_\mu\partial^\mu + m^2)\phi = bm^{-1}\gamma^{-1}\delta^3(\vec{q}-\vec{q}_p)
\end{empheq}
\vspace{-3pt}
and we demonstrate in Appendix~\ref{sec:particle_equation_A} that the trajectory equation approximately takes the form
\begin{empheq}[box=\fbox]{equation}\label{eq:equations_am2}
	d_t\left(m\gamma\vec{u}\right) = bm\gamma^{-1}\nabla\ovl{\phi}
\end{empheq}
This approximation removes the mass coupling term present in the general equations (\ref{eq:oureqs}), obviating the need to compute the continuous component $\phi\mapsto\ovl{\phi}$. In Appendix~\ref{sec:particle_equation_B}, we present an alternative derivation, using an exact (albeit non-conservative) version of the Euler--Lagrange equations (\ref{eq:lem1eqs}). Notably, the non-conservative component vanishes in the non-relativistic limit of the theory, and only affects the total energy budget of the joint particle-wave system. 

\begin{figure}[H]
	\begin{adjustwidth}{-\extralength}{0cm}
		\centering
		\begin{subfigure}{.33\linewidth}
            \centering
			\includegraphics[scale=0.33]{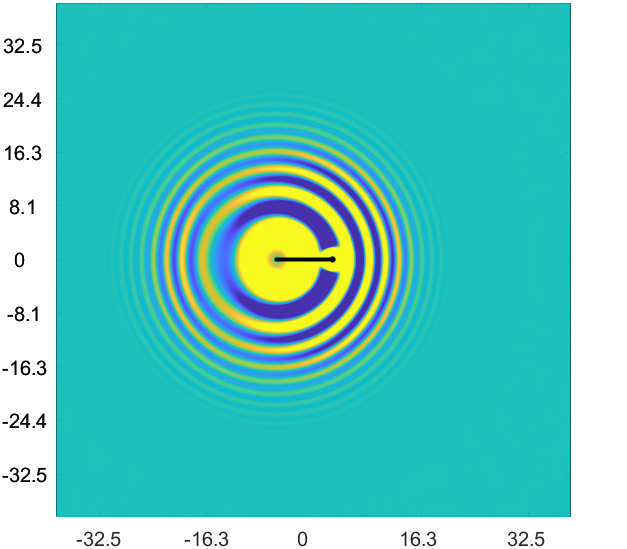}
			
            \caption{\centering}\label{fig:radiation_ex_A}
		\end{subfigure}%
		\begin{subfigure}{.33\linewidth}
            \centering
			\includegraphics[scale=0.33]{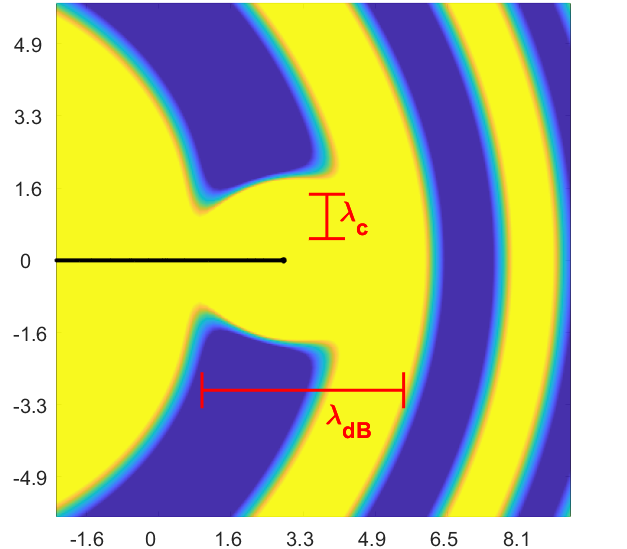}
			
            \caption{\centering}\label{fig:radiation_ex_B}
		\end{subfigure}%
        \begin{subfigure}{.33\linewidth}
            \centering
			\includegraphics[scale=0.33]{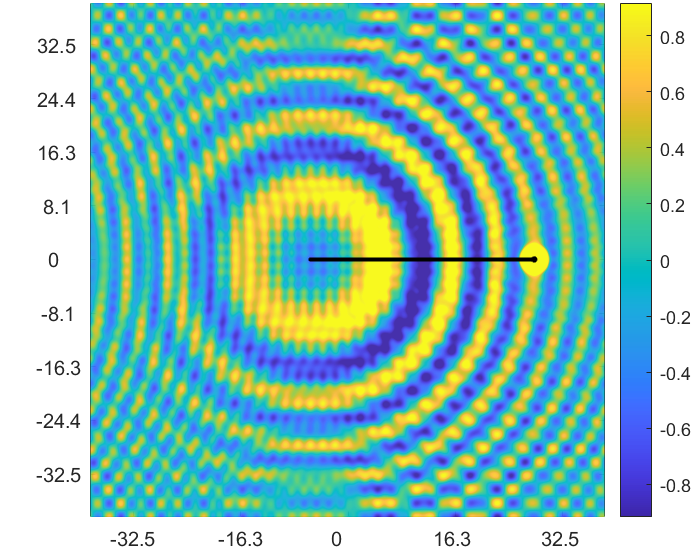}
			
            \caption{\centering}\label{fig:radiation_ex_C}
		\end{subfigure}%
	\end{adjustwidth}
	\caption{(\textbf{a}) A depiction of the free particle in our system, in two dimensions and with the coupling constant $b = 53.3$. In this simulation, we accelerate the particle from rest to a velocity $u_0 = 0.35c$. Radiation is emitted from the point of acceleration, corresponding to the form predicted in Section~\ref{sec:radiation}. Axes are given in units of $\lambda_c$. (\textbf{b}) The wave form adjoining the particle, as predicted in Section~\ref{sec:steadystate}, is visible as a high-amplitude region around the particle, of characteristic radius corresponding to the Compton scale $\lambda_c = 2\pi/m$. As predicted in Section~\ref{sec:radiation}, the local wave field has a characteristic wavelength $\lambda_{dB} = 2\pi/(\gamma m v)$, where $v$ is the \emph{instantaneous} (rather than initial) speed of the particle. (\textbf{c}) The same simulation at a later time. Because the wave travels out from the particle's point of origin, the local curvature of the wavefront decreases as the particle moves forward. Because the domain is periodic in both directions, the wave field grows complex, and the particle experiences the radiation from its periodic images.}\label{fig:radiation_ex}
\end{figure}

The form of the pilot-wave accompanying the free particle is depicted in Figure \ref{fig:radiation_ex}, albeit in two, rather than three, dimensions.
Here, the particle was accelerated from rest to a speed $u_0 = 0.35c$. We highlight four key features of the subsequent dynamics:

\begin{enumerate}
    \item The particle carries with it a high-amplitude, constant \emph{wavepacket} of radius $\sim\lambda_c$, corresponding to the bright region adjoining the particle in Figure \ref{fig:radiation_ex}. We derive the form of this wavepacket in 3D in Section~\ref{sec:steadystate}, and explore how it affects the particle's inertial mass in Section~\ref{sec:virtual};
    \item Beyond the energy imparted to the local wavepacket, energy is radiated outward from the point of acceleration in a long wavetrain. In the region around the particle, this wave always satisfies $p = \hbar k$, where $p$ is the instantaneous particle four-momentum. We derive this result in Section~\ref{sec:radiation};
    \item 
    Surfing over the underlying quasi-monochromatic wave causes the particle to vibrate in-line at the frequency $\gamma^{-1}\omega_c$, in a dynamical version of the \emph{Zitterbewegung} seen in the Dirac theory of the electron \cite{1990FoPh...20.1213H}. We examine these oscillations in Section~\ref{sec:oscillation};
    \item All of these dynamics are governed by universal balance laws for stress-energy and angular momentum, which in turn arise from our Lagrangian framework (\ref{eq:action}). We investigate the conservation of these Noether currents in Section~\ref{sec:AMconservation}, both in a general pilot-wave system and in the particular limit of interest.
\end{enumerate}

In the remainder of the paper, we rationalize each of these four features in turn, using both analytical and numerical tools. 
Finally, we combine them all in Section~\ref{sec:uncertainty} to derive 
a classical \emph{Heisenberg uncertainty principle}:
\[\sigma_x\sigma_p = K\hbar /2,\]
where $\sigma_x$ and $\sigma_p$ are variations in position and momentum in any one direction, and $K = K(b)$. We demonstrate that, for sufficiently large $b$, this reduces to the uncertainty principle of quantum mechanics.

\vspace{-6pt}

\subsection{Conserved Currents in the AM System}\label{sec:AMconservation}

We now consider the non-conservative derivation of the AM system detailed in \linebreak  Appendix~\ref{sec:particle_equation_A}. Although it does not fit into our general framework (\ref{eq:action}), it allows us to recover two key conservation laws.
The first and most important of these conservation laws is that of the \emph{stress-energy tensor} $T_\beta^{\alpha}$, which encodes system energy and linear momentum in a Lorentz-covariant 2-tensor. In a given reference frame, $E:=T^{0}_0$ can be identified with the system energy density, and similarly $p_k:=T^{0}_k$ with the momentum density. The remaining terms represent fluxes of these quantities, such that, in the absence of sources or sinks, \linebreak  we have
\[\partial_t E +\partial_k T^{k}_0 = 0,\qquad \partial_t p_i +\partial_k T_i^{ k} = 0.\]

We are able to recover a stress-energy conservation law by applying Lemma \ref{lem:noether} to spacetime translations. The full derivation is in Appendix~\ref{sec:variationproofs_B}. We define the stress-energy tensor
\begin{equation}\label{eq:SE}
	T_\alpha^\mu = -\tfrac{1}{2}\delta^\mu_\alpha(\partial_\nu\phi\partial^\nu\phi - m^2\phi^2) + \partial^\mu\phi\partial_\alpha\phi+ \delta^3(\vec{q}-\vec{q}_p)m\gamma u^\mu u_\alpha,
\end{equation}
which is exactly the sum of a relativistic free particle and a free scalar field. Then we recover the balance equations
\vspace{3pt}
\begin{empheq}[box=\fbox]{equation}\label{eq:SEcontinuity}
    \begin{gathered}
        \partial_\mu T^\mu_0 = -\delta^3(\vec{q}-\vec{q}_p) mb\gamma^{-1}u^\mu\partial_\mu\phi\\
        \partial_\mu T^\mu_k = 0 .
    \end{gathered}
\end{empheq}
This demonstrates the key benefit of our non-conservative derivation: the momentum conservation is exact, and the energy balance is neatly encoded by the material derivative $d_t\phi(q_p) := u^\mu\partial_\mu\phi(q_p)$ of the field along the particle trajectory. 

We can find a similar conservation law for the \emph{relativistic angular momentum} $M^{\sigma\nu}$ by examining spatial rotations and Lorentz boosts. Here, the spatial components $M^{k\ell}$ form the classical angular momentum bivector:
\[M^{k\ell} = q^kp^\ell - q^\ell p^k = \vec{q}\wedge\vec{p},\]
where $p^k = \tilde{T}^{k0}$ is the linear momentum of (\ref{eq:SE}). The temporal components $M^{0k} = -M^{k0}$ are somewhat less useful; we have
\[M^{0k} = q^k p^0 - q^0p^k = q^k E - tp^k,\]
which represents a scaled centre-of-mass value.

Instead of applying Noether's theorem, we deduce this conservation law more easily by leveraging (\ref{eq:SEcontinuity}). We define the angular momentum currents
\begin{equation*}
	M^{\sigma\nu\mu} = q^\sigma T^{\nu\mu} - q^\nu T^{\sigma\mu},
\end{equation*}
which gives the angular momentum $M^{\sigma\nu} = M^{\sigma\nu 0}$. 
Plugging in the balance equation (\ref{eq:SEcontinuity}) and using the symmetry of $T^{\mu\nu}$, we recover
\begin{equation*}
	\partial_\mu M^{\sigma\nu\mu} = \delta^3(\vec{q}-\vec{q}_p)(q^\sigma\partial^\nu\tau - q^\nu\partial^\sigma\tau) mbu^\mu\partial_\mu\phi(q_p),
\end{equation*}
where $\partial^0\tau=\gamma^{-1}$ and $\partial^k\tau = 0$ for $k\neq 0$.
For the spatial components $M^{k\ell\mu}$, which represent the classical angular momentum current, we deduce a true conservation law
\vspace{-3pt}
\begin{empheq}[box=\fbox]{equation*}
	\partial_\mu M^{k\ell\mu} = 0
\end{empheq}
We expect angular momentum conservation to play an important role in bound or orbiting states, which are outside the scope of this work. In the context of the linear acceleration of the free particle, this conservation law requires that the outgoing radiation have vanishing total angular momentum.

\subsection{Steady States of the Free Particle, and the Local Wavepacket}\label{sec:steadystate}
As a first step towards identifying the local form of the wave field, note that one steady-state solution of (\ref{eq:equations_am1}) and (\ref{eq:equations_am2}) is
\vspace{-6pt}
\begin{equation}\label{eq:steadystate}
	\vec{u}=0,\qquad \phi = \frac{b}{4\pi m|\vec{q}|}e^{-m|\vec{q}|}.
\end{equation}
This corresponds to a Yukawa potential of range $1/m$ or, in dimensional form, the Compton wavelength $\lambda_c$. 

Now consider a general trajectory $\vec{q}_p(t)$, and recall the (position-space) Green's function for the forced Klein--Gordon equation \cite{wolfram_greens}:
\vspace{-3pt}
	\begin{align}\label{eq:greensfn}
		\begin{split}
			G(\vec{q},t) 
			&= \frac{\theta(t)}{4\pi|\vec{q}|}\delta(t-|\vec{q}|)-\frac{m^2}{2\pi}\theta(t-|\vec{q}|)\frac{J_1(m\sqrt{t^2-|\vec{q}|^2})}{m\sqrt{t^2-|\vec{q}|^2}},
		\end{split}
	\end{align}
with $\theta$ the Heaviside step function and $J_1$ a Bessel function of the first kind. {We integrate this expression over the particle trajectory one term at a time, first defining
\[\tilde{\phi}(t_0,\vec{q}) := \int_{-\infty}^{t_0}dt\;\frac{b\gamma^{-1}}{4\pi m|\vec{q}-\vec{q}_p(t)|}\delta(t-|\vec{q}-\vec{q}_p|).\]
This term reduces to
\[\tilde{\phi}(t=0,\vec{q})=\sum\nolimits_t\frac{b\gamma^{-1}(t)}{4\pi m|\vec{q}-\vec{q}_p(t)|},\]
where the sum is taken over times $t<0$ such that $|\vec{q}-\vec{q}_p(t)|=|t|$.} Since the particle is traveling strictly slower than $c\equiv 1$, however, it can only cross this locus once---say, at $t(\vec{q})$---and the sum reduces further to
\begin{equation}\label{eq:perturbedthing}
	\tilde{\phi}(t=0,\vec{q})=\frac{b\gamma^{-1}(t(\vec{q}))}{4\pi m|\vec{q}-\vec{q}_p(t(\vec{q}))|}.
\end{equation}
Suppose we are in the instantaneous rest frame of the particle, so that $|\vec{q}_p(t)| = O(at^2)$ for some acceleration amplitude $a>0$. {For $\vec{q}$ within the ball $B_\eps(0)$, $\eps>0$, we find that $|\vec{q}_p(t(\vec{q}))| = O(a\eps^2)$, and thus $|t(\vec{q})|=|\vec{q}|+O(a\eps^2)$, yielding}
\[\tilde{\phi}(t=0,\vec{q})=\frac{1+O(a\eps^2)}{4\pi m|\vec{q}|} = \frac{1}{4\pi m|\vec{q}|}+O(a\eps).\]
With this result in hand, we reduce the total field to the local expression (\ref{eq:steadystate}); using the expression (\ref{eq:greensfn}) and subtracting the components giving (\ref{eq:steadystate}), we find
	\begin{adjustwidth}{-\extralength}{0cm}
	\begin{equation}\label{eq:radrate}
        \begin{aligned}
		\phi(t=0,\vec{q}) &= \frac{b}{4\pi m|\vec{q}|}e^{-m|\vec{q}|} - \frac{mb}{2\pi}\int_{-\eps}^0dt\;\left( \frac{J_1(m\sqrt{t^2-|\vec{q}-\vec{q}_p|^2})}{m\sqrt{t^2-|\vec{q}-\vec{q}_p|^2}}-\frac{J_1(m\sqrt{t^2-|\vec{q}|^2})}{m\sqrt{t^2-|\vec{q}|^2}}\right) + O(a\eps)\\
		&= \frac{b}{4\pi m|\vec{q}|}e^{-m|\vec{q}|} - \frac{m b}{4\pi}\int_{-\eps}^0dt\;\left( \frac{m\sqrt{t^2-|\vec{q}-\vec{q}_p|^2}}{m\sqrt{t^2-|\vec{q}-\vec{q}_p|^2}}-\frac{m\sqrt{t^2-|\vec{q}|^2}}{m\sqrt{t^2-|\vec{q}|^2}}\right) + O(a\eps)\\
		&=\frac{b}{4\pi m|\vec{q}|}e^{-m|\vec{q}|} + O(a\eps),
	\end{aligned}
    \end{equation}
	\end{adjustwidth}
using the expansion $J_1(x)=\frac{1}{2}x + O(x^2)$. In particular, the expression (\ref{eq:steadystate}) holds in a neighborhood of the particle in the particle's instantaneous frame of reference, up to a finite contribution of amplitude $O(a\eps)$.
By performing a Lorentz boost in the $\hat{e}_1$ direction, we recover the more general form
\begin{equation}\label{eq:steadystategeneral}
	\phi = \frac{b\exp\left(-m\sqrt{\gamma^2(q^1-tu^1)^2+(q^2)^2+(q^3)^2}\right)}{4\pi m\sqrt{\gamma^2(q^1-tu^1)^2+(q^2)^2+(q^3)^2}}
\end{equation}
for the wavepacket adjoining a particle of velocity $\vec{u} = u^1\hat{e}_1$. As we discuss further in Appendix~\ref{sec:covariance}, this extension follows exactly from the approximate Lorentz-covariance of the AM system. 

In short, the analysis above demonstrates that the particle is \emph{dressed with a trajectory-independent Yukawa potential}, constant up to a length contraction. In Section~\ref{sec:virtual}, we will demonstrate that this ``wavepacket'' modifies the particle's effective inertial mass and momentum. We see a numerical depiction of this wavepacket in Figure \ref{fig:radiation_ex}, albeit in two rather than three dimensions. There, the local wavepacket corresponds to the high-amplitude region of radius $\sim\lambda_c$ centered on the particle.

Another important inference may be made by re-examining (\ref{eq:radrate}). Consider again the rest frame of the particle at $t=0$, and suppose as before that the particle is accelerating as $|\vec{q}_p(t)| = O(at^2)$. Then the second term in (\ref{eq:radrate}) corresponds precisely to the \emph{radiation} created by the particle's motion between time $t=-\eps$ and $t=0$; that is, the change in the scalar value of the field $\phi$ away from its steady state (\ref{eq:steadystategeneral}). Our calculation shows that this value is at most $O(a\eps) = O(\Delta v)$, where $\Delta v = O(a\eps)$ is the magnitude of the velocity change in this interval. 
If the particle is not accelerating at all (and so $a=0$), it does not radiate any waves outward: at a fixed velocity $\vec{u}$, the particle carries \emph{only} the wavepacket (\ref{eq:steadystategeneral}). If the particle acceleration has a magnitude $O(a)$, as in the above analysis, the value of $\phi$ at a later time is changed at most \emph{at the rate} $O(a)$. The resulting radiation becomes significant only if $a\eps = O(1)$, as will arise if the particle rapidly changes from one velocity state to another. This calculation will be substantiated in our numerical study of wave radiation in Section~\ref{sec:radiation}.

\subsection{Zitterbewegung: Particle Oscillation at the Compton Frequency}\label{sec:oscillation}


Spontaneous particle oscillations have been shown to arise in several models of classical 
pilot-wave systems. In-line speed oscillations have been reported to arise in several 
settings in the walking droplet systems, including the hydrodynamic 
analogue of Friedel oscillations~\cite{Saenz2019a}.
In-line oscillations with amplitude comparable to the wavelength of the pilot wave have been shown 
to be a robust feature of the \emph{generalized pilot-wave framework} (GPWF)~\cite{Durey2021}, a parametric generalization of the walking droplet system~\cite{Bush2015a}. Moreover, one-dimensional motion of the free particle in 
HQFT~\cite{Dagan2020,Durey2020b} is marked by erratic in-line oscillations at the Compton frequency.
Notably, in all of these examples, particle oscillations are restricted to the in-line direction, and the {coupling strength between particle and wave is a \emph{periodic function of time}}. 

We proceed by demonstrating that both features of de Broglie's 
harmony of phases, an internal particle oscillation at frequency $\gamma^{-1}\omega_c$ and an accompanying wave of wavelength $\lambda_{\text{dB}}$, emerge naturally from the time-invariant dynamics (\ref{eq:equations_am1}) and  (\ref{eq:equations_am2}).
Moreover, the oscillation frequency and wavelength update dynamically as the particle's momentum changes, in order to preserve the de Broglie relation $p=\hbar k$. Finally, the particle vibrates in \emph{all} directions when interacting with a wall-bounded geometry.


In our first series of tests, we start a particle at rest and accelerate it quickly to a speed $u_0$, from which it settles quickly into a steady speed $u=u(u_0,b)<u_0$. We discuss the form of $u(u_0)$ further in Section~\ref{sec:virtual}, and in particular, we derive a nonzero \emph{virtual mass} imparted to the particle's rest mass by the surrounding wavefield.

\begin{figure}[H]
		\includegraphics[scale=0.3]{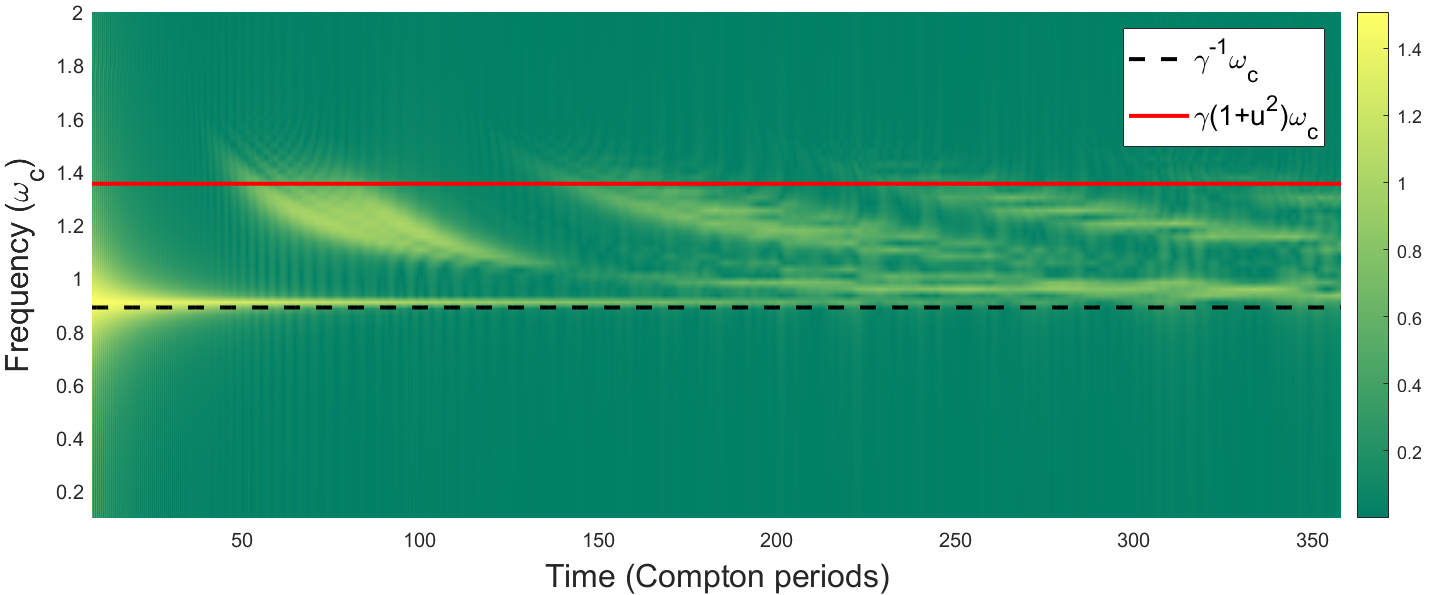}
		\caption{A spectrogram of in-line oscillations for our two-dimensional system, with coupling constant $b = 53.3$. {The shown color values  are normalized by $\Delta x\mapsto\arctan(50\cdot\Delta x/\lambda_c)$, where $\Delta x$ is the oscillation magnitude.} Here, we give the particle an initial velocity $u_0/c=0.5c$, which quickly relaxes to a mean velocity $u/c=0.455c$. For $t < 50$ Compton periods, the particle undergoes an oscillation at the frequency $\gamma^{-1}\omega_c$. Thereafter, waves cover the entire periodic domain, and the particle vibrates at frequencies between $\gamma^{-1}\omega_c$ and $\gamma(1+v^2)\omega_c$. Note, the diminishing intensity of the yellow line at $\gamma^{-1}\omega_c$ reflects the temporal decay of the in-line Zitter.
  }\label{fig:spectrogram}
\end{figure}

For $u_0 = 0.5c$ and $b=53.3$, a spectrogram of the resulting in-line position oscillations is shown in Figure \ref{fig:spectrogram}. We highlight two noteworthy effects:
\begin{enumerate}
	\item Initially, the particle undergoes in-line oscillations with frequency $\gamma^{-1}\omega_c$ in response to outgoing radiation from the point of acceleration, effectively surfing over its own radiative wave field. The form of this radiation is discussed in Section~\ref{sec:radiation};
	\item After $\sim 50$ Compton periods, the particle oscillates with amplitude $n_b^{-1}\gamma^{-e_b}\lambda_c\sim \lambda_c$ at frequencies between $\gamma^{-1}\omega_c$ and $\gamma(1+u^2)\omega_c$. This is an artifact of our periodic domain, and specifically the particle interacting with the wave form generated by its periodic images. However, we expect the same effect to occur any time the particle interacts with a wall-bounded geometry, and its wave form reflects off the boundaries.
	\vspace{-18pt}
\end{enumerate}

\begin{figure}[H]
	\begin{adjustwidth}{-\extralength}{0cm}
		\centering
		\begin{subfigure}{.4\linewidth}
			\centering
			\includegraphics[scale=0.32,valign=c]{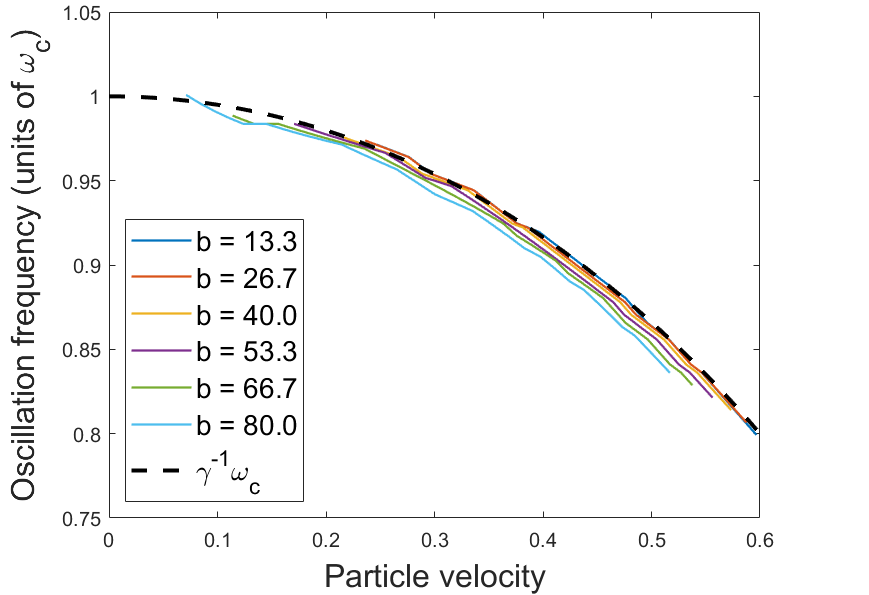}
			
            \caption{\centering}\label{fig:oscillations_A}
			
		\end{subfigure}%
		\begin{subfigure}{.4\linewidth}
			\centering
			\includegraphics[scale=0.32,valign=c]{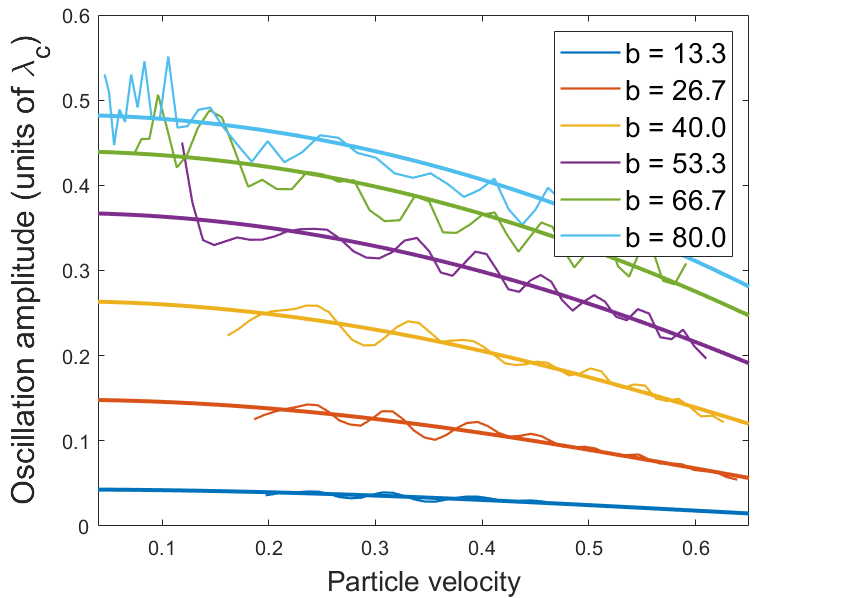}
			
            \caption{\centering}\label{fig:oscillations_B}
		\end{subfigure}%
		\begin{subfigure}{.2\linewidth}
			\centering
            \adjustbox{valign=c}{
			\begin{tabular}{c cc}
				\toprule
				$b$ & $e_b$ & $n_b$\\\midrule
				$13.3$ & $3.88$ & $238.$\\
				$26.7$ & $3.52$& $69.$\\
				$40.0$ & $2.87$& $39.$\\
				$53.3$ & $2.38$& $28.$\\
				$66.7$ & $2.10$& $23.$\\
				$80.0$ & $1.97$& $21.$\\\bottomrule
				
			\end{tabular}
            }
            \vspace{0.4in}
            \caption{\centering}\label{fig:oscillations_C}
		\end{subfigure}
	\end{adjustwidth}
	
	\caption{(\textbf{a}) Dominant oscillation frequencies at the beginning of each trajectory, for particles across the range of initial velocities $u_0=0.2$ to $u_0=0.65$. We observe that the particle oscillates at the frequency $\gamma^{-1}\omega_c$ independent of the coupling constant and velocity.
	(\textbf{b}) Amplitudes of in-line oscillations in the long-time limit of Figure \ref{fig:spectrogram}, i.e., after waves have covered the entire periodic domain. Curves of the form $n_b^{-1}\gamma^{-e_b}\lambda_c$ are shown for reference, where $n_b$ and $e_b$ are least-squares fits (reported in the Table~(\textbf{c})).}
      \label{fig:oscillations}
\end{figure}

We quantify the first effect in Figure \ref{fig:oscillations}a, where we repeat this experiment across a wide range of velocities and $b$ values. We see that the oscillation frequency conforms closely to $\gamma^{-1}\omega_c$ in all cases. In all simulations, the dominant frequency is constant until the particle encounters radiation from the opposite side of the domain.

\vspace{0pt}

\begin{mdframed}
	\paragraph*{\textbf{Numerical Simulation of the AM System.}}
	Our numerical code is based on that developed by Faria \cite{Faria_2016} for walking droplets. We model the space as a two-dimensional periodic domain, allowing us to use high-accuracy pseudo-spectral methods to resolve the wave field. 
	In turn, we evolve the wave field using a fourth-order Runge--Kutta method, separately tracking $\phi$ and $\eta:=\partial_t\phi$ in order to break (\ref{eq:oureqs}) into two first-order equations.
	
    Our algorithm takes the following form. At each timestep, we add an approximate delta function---taken to be a Gaussian of variance $2m^{-2}=2\lambda_c^2$---to the field $\eta$ at the current particle position $\vec{q}_p$, scaled by $b\gamma^{-1}$. Suppose $\phi_{\vec{\ell}}\;$ and $\eta_{\vec{\ell}}\;$ are the discrete Fourier transforms of $\phi$ and $\eta$, respectively, so that, for instance, $\phi = \sum \phi_{\vec{\ell}}\;e^{i\vec{\ell}\cdot\vec{q}}$. We evolve the fields $\phi$ and $\eta$ for one time-step according to the equations
	\[\partial_t\phi_{\vec{\ell}}\; = \eta_{\vec{\ell}}\;,\qquad\partial_t\eta_{\vec{\ell}}\; = (-|\ell|^2+m^2)\phi_{\vec{\ell}}\;,\]
	and calculate the gradient $\nabla\phi(\vec{q}_p)$ using the field's Fourier expansion:
	\[\nabla\phi(\vec{q}_p) = \sum i\vec{\ell}\phi_{\vec{\ell}}\;e^{i\vec{\ell}\cdot\vec{q}_p}.\]
	We use this to evolve $\gamma \vec{u}$ for one time-step, and we move the particle accordingly.
	
\end{mdframed}

As we argue in Appendix~\ref{sec:covariance}, we can deduce the same behavior after a Lorentz transformation. Namely, oscillations at $\gamma^{-1} \omega_c$ arise regardless of initial velocity. If the particle is moving at a velocity $\vec{u}'$ and accelerates to a steady velocity $\vec{u}$, it begins to  vibrate at the frequency $\gamma_{u}^{-1}\omega_c$. This demonstrates a relativistically-correct internal clock at the Compton frequency, as in de Broglie's original model, but with two distinctions. First, it emerges naturally from a time-invariant dynamics, without reference to intrinsic particle oscillation. 
Second, the emergent particle clock updates dynamically to adjust to the particle's \emph{current} velocity, preserving the synchrony of de Broglie's clock even under the influence of applied forces. We detail the wave form complementing this clock in the subsequent section.

\subsection{A Dynamical Harmony of Phases}\label{sec:radiation}

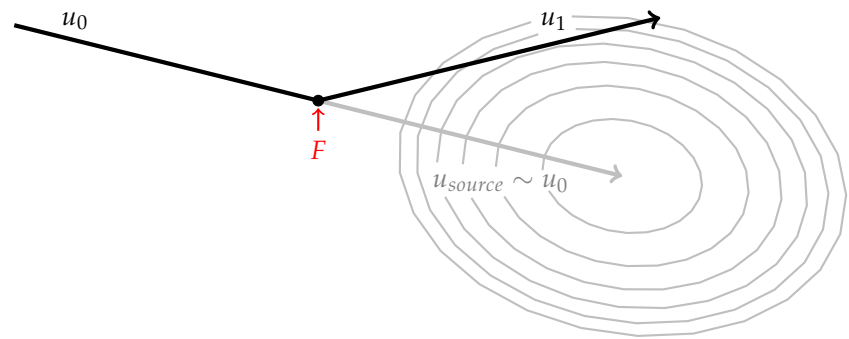
\begin{figure}[H]
		\begin{tikzpicture}
        \foreach \i in {6,...,7}
        {
                \draw [lightgray,thick,domain=122:471] plot ({5 + 1.5*ln(\i)*cos(\x)+0.7*0.3*ln(\i)*sin(\x)}, {-1 + 0.7*1.5*ln(\i)*sin(\x)
             - 0.3*ln(\i)*cos(\x)});
        }
        \foreach \i in {2,...,5}
        {
                \draw [lightgray,thick,domain=(-145-2*\i):190] plot ({5 + 1.5*ln(\i)*cos(\x)+0.7*0.3*ln(\i)*sin(\x)}, {-1 + 0.7*1.5*ln(\i)*sin(\x)
             - 0.3*ln(\i)*cos(\x)});
        }
        \draw[lightgray, ultra thick, ->] (1,0) -- (5,-1);
        \draw[black, ultra thick, ->] (-3,1) -- (1,0) -- (5.5, 1.1);
        \filldraw[black] (1,0) circle (2pt) ;
        \draw[red, thick, ->] (1,-0.4) -- (1,-0.1) ;
        \filldraw[red] (1,-0.4) circle (0pt) node[anchor=north]{$F$};
        \filldraw[black] (-2.2,1.3) circle (0pt) node[anchor=north]{$u_0$};
        \filldraw[black] (4.1,1.3) circle (0pt) node[anchor=north]{$u_1$};
        %
        \filldraw[gray] (3.4,-0.85) circle (0pt) node[anchor=north]{$u_{source}\sim u_0$};
        \end{tikzpicture}
    \caption{A diagrammatic sketch of the radiation pattern following a general acceleration of the particle in our system. 
    Suppose the particle (shown in black) initially moves at a steady velocity $u_0$, carrying with it the Yukawa wavepacket (\ref{eq:steadystategeneral}), and is then accelerated by an external force $F$ to a new velocity $u_1$. This acceleration spawns a new, continuous source of waves that drifts along the \emph{extrapolated} original trajectory of the particle (shown in gray). The wave source spreads out in the ellipsoid (\ref{eq:wavefrontellipse}) at a speed $u_\text{expansion}\sim|\vec{u}_1 - \vec{u}_0|$ given by (\ref{eq:wavefrontexpansion}), all the while drifting at the velocity $\vec{u}_\text{source}\sim\vec{u}_0$ defined by (\ref{eq:wavefrontvelocity}). We stress that the gray curves are approximate level sets of the wave amplitude, but not generally of the wave phase or wavenumber.
    }\label{fig:diagram}
\end{figure}

We proceed to examine the wave form generated following a particle acceleration, first for the case of acceleration from rest (see Figure \ref{fig:radiation_ex}), then for a more general acceleration (see Figure \ref{fig:diagram}). We shall demonstrate that the resulting wave form naturally gives rise to the periodic forcing needed for the Zitterbewegung reported in the previous section. 
Specifically, we will show that the oscillation frequencies reported in Section~\ref{sec:oscillation} arise from the particle being repeatedly washed over by quasi-monochromatic waves of wavelength $\lambda_{dB}$ 
and phase velocity $c^2/u$.
This analysis allows for a detailed comparison of the wave forms arising in pilot-wave hydrodynamics, in HQFT, and in our new model.

We first review some elementary properties of Klein--Gordon waves before returning to our particular system. Consider the dispersion relation of Klein--Gordon waves:
\[\omega^2 = m^2 + |\vec{k}|^2,\]
or in dimensional form, $\omega^2 = \omega_c^2 + c^2|\vec{k}|^2$.
Here, $\omega$ is the local oscillation frequency and $\vec{k}$ the local wavevector. The group velocity of the wave is
\begin{equation}\label{eq:groupvelocity}
    \vec{v}_g = \nabla_{\vec{k}} \omega = \frac{\vec{k}}{\sqrt{m^2 + |\vec{k}|^2}},
\end{equation}
which corresponds precisely to the velocity of a point-particle of mass $m$ and momentum $\vec{k}$, as can be seen by inverting the above relationship: $\vec{k} = m\gamma_{v_g}\vec{v}_g$.
The system is dispersive: the group velocity depends on wavelength. Thus, following a wave disturbance from a point source, different component wavelengths $\lambda := 2\pi/k$ travel outwards at different speeds $v_g = k/\sqrt{m^2+k^2}$.
The result is a wave train centered on the original source, with each excited wavenumber $k$ spreading outward at its corresponding group speed $v_g$. If a particle of mass $m$ is traveling away from this point-source at a velocity $\vec{u}$, the region immediately adjoining the particle must then carry a wavelength $\lambda_{dB} = 2\pi\hbar/|\vec{p}|$, where $\vec{p} = \gamma m\vec{u}$ is the particle's momentum. This region grows in size over time---as wave crests of wavenumber $\vec{k}$ and $(\vec{k}+\eps)$ drift apart with velocity $O(\eps/k)$---giving rise to a quasi-monochromatic wave form in the vicinity of the particle. 

The above behavior occurs in our system when a particle is accelerated from rest (see Figure \ref{fig:radiation_ex}).
At the particle's point of acceleration---the origin, in the case of Figure \ref{fig:radiation_ex}---the particle excites a range of wavevectors $\vec{k}$ defined by
\begin{equation}\label{eq:waveball}
    \hbar|\vec{k}| \lesssim m\gamma u,
\end{equation}
where $u$ is the new particle speed. The different components spread out from the origin in a spherical wave train according to (\ref{eq:groupvelocity}), and the particle surfs along the outgoing, origin-centered, spherical region of local wavelength $\lambda_{dB}$. The particle is repeatedly washed over by these waves, which possess a phase speed 
\[\omega/k = \sqrt{k_{dB}^2+m^2}/k_{dB} = 1/u = c^2/u.\]
The relative speed between the wave phase and the particle is then $c^2u^{-1}-u$. We may thus deduce the forcing frequency of the 
wave 
on the particle by multiplying the relative velocity by the local wavenumber:
\[\omega_\text{particle} = \left(u^{-1}-u\right)k_{dB} = \left(u^{-1}-u\right)\gamma m u = \gamma^{-1}\omega_c.\]
This result is in accord with the estimate $\omega_\text{particle}=\omega_c\gamma^{-1}$ evident in Figure \ref{fig:oscillations} for $t<50$ Compton periods, that is, before the particle encounters radiation from its periodic image sources. We note that the response to the quasi-monochromatic waves of wavelength $\lambda_{dB}$ in the particle's immediate vicinity is the dominant effect on the subsequent particle motion. In Section~\ref{sec:steadystate}, we have shown that the particle emits radiation at a rate proportional to its instantaneous acceleration. As such, waves generated by the particle's resulting in-line Zitter over a single oscillation are at most of amplitude $O(\eps)$, where $\eps \lambda_c$ is the amplitude of particle vibration. We plot values of $\eps=n_b^{-1}\gamma^{-e_b}$ in Figure \ref{fig:oscillations}b, but note that $\eps\ll 1$ in all cases. Finally, we note that in an unbounded domain, the resulting in-line Zitter dies down over time, since the wave amplitude necessarily decays as the wave disperses. This behavior may be seen by comparing Figure \ref{fig:radiation_ex}a--c.



Upon accelerating, the particle excites a range of wavenumbers \emph{including} the set (\ref{eq:waveball}). We proceed to demonstrate that the dominant wavenumbers of the outgoing wavefront are, in fact, bounded above in norm by the value $\sim k_{dB}$. Consider the behavior of the particle for $t>50$ Compton periods in Figure \ref{fig:spectrogram}, when the particle
has circled the far side of its periodic domain and 
encounters its own waves \emph{head-on}. {The relative speed between the particle and \emph{oncoming} de Broglie waves is now $c^2u^{-1}+u$, so, following a similar derivation as above, the excited vibration frequency in the particle is}
\[\omega_\text{max} = \gamma(1+u^2)\omega_c.\]
{In Figure \ref{fig:spectrogram}, this value} appears as an approximate upper-bound for the particle oscillations; {higher oscillation frequencies, which would correspond to waves of smaller wavelength,} 
are not evident. {Although we see in Figure \ref{fig:radiation_ex}a that some waves of momentum $k>m\gamma u$ are excited---i.e., those moving faster than the particle itself---}
this spectrogram demonstrates that they are negligible compared to their lower-momentum counterparts.
Thus, in accelerating from rest
, the particle effectively excites only the wavevectors (\ref{eq:waveball}).



We emphasize that
when the particle starts from rest, waves are radiated outward \emph{from the point of acceleration}. This marks a point of contrast with the behavior in the walking droplet system~\cite{Couder2005a} and HQFT~\cite{Dagan2020,Durey2020b}, where waves are radiated continuously along the particle's trajectory. In our system, the free particle travels alongside a nearly-planar wavefront of the de Broglie wavelength, as shown in Figure \ref{fig:radiation_ex}. Finally, our results support the prediction of Section~\ref{sec:steadystate}, that a nearly constant-velocity state of the free particle is also nearly non-radiating; specifically, an $O(\eps\lambda_c)$-amplitude vibration about a constant velocity induces only an $O(\eps\omega_c)$ rate of radiation. 


We now consider a more general particle acceleration: a particle moving steadily with velocity $\vec{u}_0$ is accelerated quickly to a new mean velocity $\vec{u}_1$.
As we show rigorously in Appendix~\ref{sec:particle_equation_B}, we can apply a Lorentz transformation to reduce this case to that treated previously, and thus deduce the form of the resulting pilot wave. 
Figure \ref{fig:diagram} illustrates the form of radiation arising in a 
Lorentz-boosted coordinate system.
The wave can no longer simply radiate from the point of acceleration, which is not a well-defined notion under Lorentz symmetry.
Instead, a new, continuous wave source is spawned at the point of acceleration, and travels along the \emph{extrapolated, original trajectory} of the particle at a velocity $\vec{u}_\text{source}$ comparable to $\vec{u}_0$, to be derived shortly. This ``virtual'' source continues to radiate waves in all directions, their form depending on the velocity \emph{difference} between the particle and wave source. 

To quantify this effect, suppose that $\vec{v}\sim \vec{u}_1 - \vec{u}_0$ is the velocity of the outgoing particle in the rest frame of the incoming particle. Without loss of generality, we suppose that $\vec{u}_0 = u_0\hat{e}_1$ lies in the $x$ direction. In the rest frame of the incoming particle, the outgoing wavefront takes the form described above: a spherically symmetric wavefront expanding (approximately) at the speed $v=|\vec{v}|$. In particular, the wavefront satisfies the equations
\[(x')^2 + (y')^2 + (z')^2 = v^2(t')^2.\]
Transforming back to the laboratory frame, the wavefront must satisfy the transformed equations
\begin{equation}\label{eq:wavefrontellipse}
    \frac{1-u_0^2v^2}{1-u_0^2}\left(x - \frac{1 - v^2}{1-u_0^2v^2} u_0t\right)^2 + y^2 + z^2 = \frac{1-u_0^2}{1-u_0^2v^2}v^2t^2.
\end{equation}
Specifically, this is the Lorentz transformation of the fastest-moving wavefront in the particle's rest frame (wherein $\lambda=\lambda_{dB}$), which approximately defines the outer edge of the full wave form. 

{One must be careful in interpreting the wave geometry in our new frame of reference. In a frame where the particle accelerates from rest, the outer edge of the wave form is a level set simultaneously of wavenumber, phase, and amplitude, but such is not the case in general. First, note that in transforming from the particle's original rest frame to our new, laboratory frame, the wavenumber becomes higher in front of the particle and lower behind it; thus, the locus (\ref{eq:wavefrontellipse}) is no longer a level set of the wavenumber}. It is also no longer a level set of the wave phase; indeed, the front (\ref{eq:wavefrontellipse}) {corresponds to values of $\phi$} from a finite time-interval in the particle's original rest frame (specifically, an interval $\sim (vu_0/c^2)t'$, following the equations of a Lorentz transform), over which the high frequency $\sim\omega_c = mc^2/\hbar$ of the wave would spread the wave front over a range of phases $\Delta\theta\sim \omega_c(vu_0/c^2)t'=(mu_0v/\hbar)t'$. However, it \emph{is} approximately a level set of the wave amplitude, which is a Lorentz scalar (avoiding the first problem) and modulates far slower than the phase (avoiding the second). 
We proceed by thinking of it in these terms.
Other amplitude level sets, corresponding to longer wavelengths in the particle's rest frame, can be found by replacing $v$ with the corresponding group speed (\ref{eq:groupvelocity}).

The relation (\ref{eq:wavefrontellipse}) reveals the wave form to be ellipsoidal, dilated in the direction $\vec{u}_0$ by a factor $\sqrt{(1-u_0^2v^2)/(1-u_0^2)}$. The center of the ellipsoid is traveling with the velocity
\begin{equation}\label{eq:wavefrontvelocity}
\vec{u}_\text{source} = \frac{1 - v^2}{1-u_0^2v^2} \vec{u}_0,
\end{equation}
which reduces to $\vec{u}_0$ if either $v\to 0$ or $u_0\to 0$. The ellipsoid is expanding at the speed
\begin{equation}\label{eq:wavefrontexpansion}u_\text{expansion} = v\sqrt{\frac{1-u_0^2}{1-u_0^2v^2}},
\end{equation}
which reduces to $v$ in the same limits. The excited wavevectors in this wave form are necessarily given by the Lorentz-transformed version of the ball (\ref{eq:waveball}) in reciprocal space.  

Figure \ref{fig:diagram} sketches level sets of the wave amplitude following a general acceleration, as discussed above, but we stress again that these are \emph{not} level sets of the phase or of the wavenumber; in fact, the phase generally oscillates rapidly around the outer edge of the wavefront, and up to a rescaling, the wavevector aligns exactly with the group velocity of the immediate wave field, including drift. A particular outcome of this is that even though the particle appears to be traveling askew (not perpendicular) to the plane of the wavefront in Figure \ref{fig:diagram}, the wavevector at the particle's position \emph{must} (by Lorentz symmetry) match the new momentum of the particle, and thus the phase increases most strongly \emph{in the particle's direction of motion}. By Lorentz symmetry, we also see that this wave form continues to drive particle oscillations at the Compton frequency, although now, these generally have both in-line and lateral components. The wave form thus continues to surround the particle with a quasi-monochromatic region of wavelength $\lambda_{dB}$, or wavenumber $k_{dB}$, even though the wavevector is \emph{not} perpendicular to the amplitude level sets depicted in Figure \ref{fig:diagram}.

Taken together with the particle oscillation of the previous section, we can understand the radiative behavior in our system as a dynamical version of de Broglie's \emph{harmony of phases}, wherein the particle's internal clock and the underlying guiding wave remain locked in phase throughout its motion. Just as the particle's vibration updates dynamically to remain at the characteristic frequency $\gamma^{-1}\omega_c$, the wavelength of the field (at the particle position) updates to preserve the relation $p=\hbar k$. Together, these effects lock the particle and wave in phase---necessarily, as the wavelength update \emph{creates} the corresponding frequency update---and preserves the harmony of phases throughout the particle motion.

\subsection{Virtual Mass in the Local Wavepacket}\label{sec:virtual}

We proceed by highlighting a critical feature of the free pilot-wave system, made evident by the local wavepacket (\ref{eq:steadystategeneral}): in steady state, the particle \emph{shares} a certain amount of energy with the field around it, spread over a radius $\sim\lambda_c$. We refer to this as a \emph{virtual mass}. While it is not readily apparent in the equations (\ref{eq:oureqs}), it impacts the particle's inertial response through a constant augmentation $m\mapsto m + \delta m$ of the particle mass. Quantitatively, this virtual mass is exactly the energy carried by the wavepacket (\ref{eq:steadystategeneral}); because the wavepacket is independent of particle dynamics, so too is the virtual mass $\delta m$.

Now, we note that the \emph{virtual mass fraction} $\delta m/m$ likely differs between our two-dimensional simulations and the three-dimensional model developed previously. With that in mind, we derive this effect in a three-dimensional system, but provide numerical results only for the two-dimensional AM system.

Recall the momentum continuity equation (\ref{eq:SEcontinuity}):
\[\partial_\mu T^\mu_k=:\partial_\mu((T_\text{field})^\mu_k + (T_\text{part.})^\mu_k) = 0,\]
with
\[(T_\text{part.})^\mu_k = \delta^3(\vec{q}-\vec{q}_p)\gamma m u^\mu u_k,\]
\[(T_\text{field})^\mu_k = -\tfrac{1}{2}m^2\delta^\mu_k(\partial_\nu\phi\partial^\nu\phi - m^2\phi^2) + m^2\partial^\mu\phi\partial_k\phi.\]
Note that these are simply the momentum and momentum flux densities of a free particle and field, respectively. In a fixed reference frame, we define the corresponding three-momenta as the space integrals
\[(P_\text{part.})_k = \int d^3\vec{q}\; (T_\text{part.})^{0}_k = \gamma mu_k,\]
\[(P_\text{field})_k = \int d^3\vec{q}\;(T_\text{field})^{0}_k = m^2\int d^3\vec{q}\;\partial_t\phi \partial_k\phi,\]
which satisfy the conservation law
\[d_t(P_\text{part.}+P_\text{field}) = 0, \]
obtained by integrating (\ref{eq:SEcontinuity}) over space.
Now, decompose the wave field as $\phi = \phi_\text{wav} + \phi_\text{rad}$, where $\phi_\text{wav}$ is simply the wavepacket (\ref{eq:steadystategeneral}). This gives us another natural field momentum,
\[(P_\text{wav})_k = \int d^3\vec{q}\;(T_\text{wav})^{0}_k := m^2\int d^3\vec{q}\;\partial_t(\phi_\text{wav})\partial_k(\phi_\text{wav}),\]
which dominates the total field momentum in the case that radiation rates are small. Such is true in all of our numerical experiments, as we quantify in Appendix~\ref{sec:particle_equation_A}.

Since the particle must bring along a wavepacket of the form $\phi_\text{wav}$, it is the \emph{combined} momentum 
\[P_\text{eff.} = P_\text{part.} + P_\text{wav}\]
that determines the particle's inertial response. The remainder $P_\text{field} - P_\text{wav}$ determines only higher-order couplings to the underlying wave field.

Since $\phi_\text{wav} = \phi_\text{wav}(\vec{q} - \vec{u} t)$, we find that $\partial_t\phi_\text{wav} = u^k\partial_k\phi_\text{wav} = \vec{u}\cdot\nabla\phi_\text{wav}$. Critically, we note that the integral for $P_\text{wav}$ only involves derivatives of $\phi_\text{wav}$ in the direction of $\vec{u}$. Indeed, the coefficient $\vec{u}\cdot\nabla\phi_\text{wav}$ only depends on this directional derivative by construction, and likewise for the vector $\nabla\phi_\text{wav}$ by symmetry. We can then write the integrand as
\begin{align*}
	P_\text{wav} = b^2\gamma^2\vec{u}\cdot\int d^3\vec{q} \; \vec{U}\otimes\vec{U},
\end{align*}
using the notation $\vec{U}:=mb^{-1}(\nabla\phi)(\vec{q})$ to emphasize that the latter is now a $\vec{u}$-independent function of space. Finally, we switch coordinates of integration to those of the rest frame of the particle, $\vec{q}'$, and thus pick up a factor of $\gamma^{-1}$:
\[P_\text{wav}=b^2\gamma\vec{u}\cdot\int d^3\vec{q}'\;\vec{U}\otimes\vec{U} =: b^2\gamma\vec{u}\cdot A.\]
Here, $A\in \mathbb{R}^3\otimes\mathbb{R}^3$ depends only on the mass density $m$ of the field. Recall from symmetry that $P_\text{wav}$ must point parallel to $\vec{u}$, restricting $A = (\delta m) \mathbf{1}$ for some $\delta m\in\mathbb{R}^3$ independent of velocity. In fact, we know that $\delta m$ must be proportional to $m$, as the only remaining mass scale in the system. In total, we find that
\begin{empheq}[box=\fbox]{equation}\label{eq:momentumfraction}
	P_\text{wav} = b^2(\delta m)\gamma\vec{u},\qquad P_\text{eff.} = (m + b^2\delta m)\gamma\vec{u} = \gamma m_\text{eff}\vec{u}
\end{empheq}
This \emph{effective} mass, rather than the ``decoupled mass'' $m$, thus determines the inertial response of the particle. 

We can confirm the influence of the wave-induced virtual mass by looking again at the numerical experiments of Section~\ref{sec:oscillation}. Recall that, in those experiments, we start a particle at rest, impart a momentum $\vec{p}_\text{tot} = \gamma m\vec{u}_0$ to the particle, and observe its evolution. Though we focused on velocity oscillations before, we now look at how the particle approaches a steady-state speed $u_s<u_0$.

\begin{figure}[H]
	\begin{adjustwidth}{-\extralength}{0cm}
		\centering
		\begin{subfigure}{.5\linewidth}
			\centering
			\includegraphics[scale=0.36]{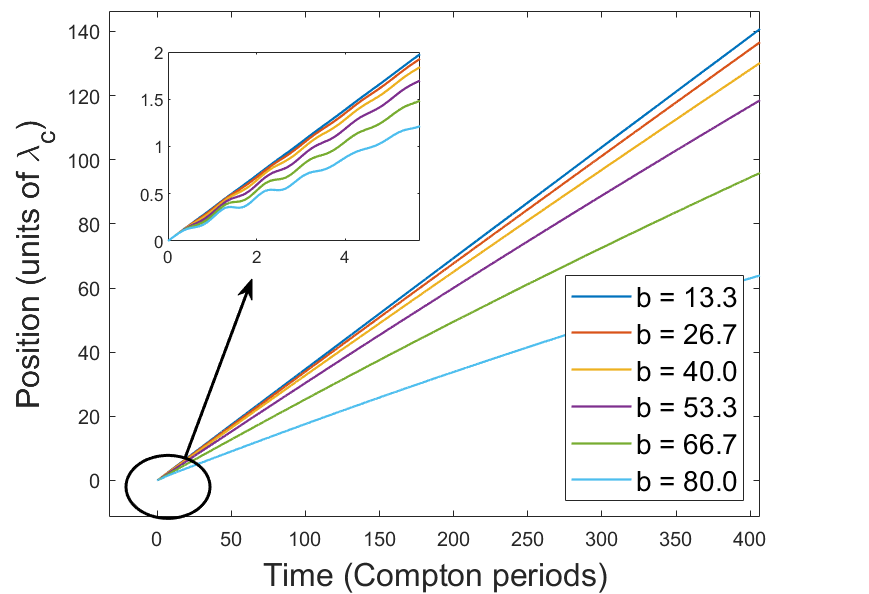}
			
            \caption{\centering}\label{fig:radiationA}
        
		\end{subfigure}%
		\begin{subfigure}{.5\linewidth}
			\centering
			\includegraphics[scale=0.36]{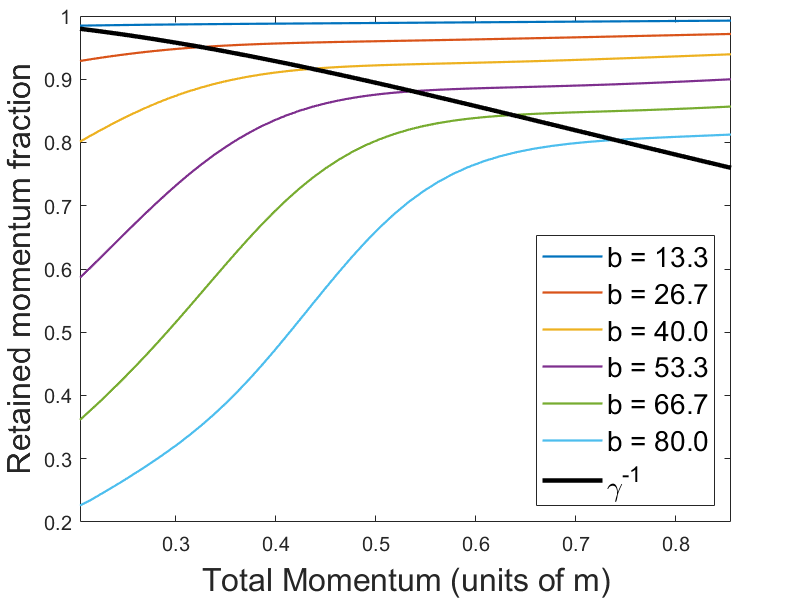}
			
			\caption{\centering}\label{fig:radiationB}
            
		\end{subfigure}
	\end{adjustwidth}
	\caption{(\textbf{a}) Six particle trajectories in our two-dimensional system, all given an initial velocity $u_0 = 0.35c$ (or $p_0 \sim 0.37mc$). Within a Compton timescale, the particle exchanges a certain amount of momentum with the underlying wave field and settles into a lower, steady-state velocity. (\textbf{b}) Fraction of total momentum retained by the particle after an initial acceleration from rest, with zero initial wave field. The resulting momentum transfer depends on both $p_0$ and $b$, and can be broken into two states. The first is the ``high-radiation regime'' below the black curve, where the particle first loses much of its momentum to the outgoing wavefront (\ref{eq:wavefrontellipse}). The second is the ``low-radiation regime'' above the black curve, where momentum is transferred almost exclusively to the local wavepacket (\ref{eq:steadystategeneral})---i.e., only the momentum $b^2(\delta m)\gamma\vec{u}$ is lost by the point particle.}\label{fig:radiation}
\end{figure}

Figure \ref{fig:radiation}a shows horizontal particle position after imparting a velocity $u_0=0.35c$. 
First, note that the relaxation to the steady-state speed $u_s<u_0$ occurs over the Compton timescale (roughly the period of several oscillations, as seen in the cutout). This short-time dynamics is characterized by a transfer of momentum from the particle to its adjoining wavepacket, as predicted by Corollary \ref{cor:exchange}. We can think of this exchange as a reflection of the particle's 
delocalised nature: since the local wavepacket requires momentum everywhere over a radius $\sim \lambda_c$, it requires a time $\sim1/\omega_c$ to distribute momentum appropriately. This also explains why radiation closely follows a single-source approximation, as we encountered in the preceding section; radiation occurs upon particle-to-wave energy transfer, which occurs over
a Compton timescale.

\begin{figure}[H]
		\includegraphics[scale=0.36]{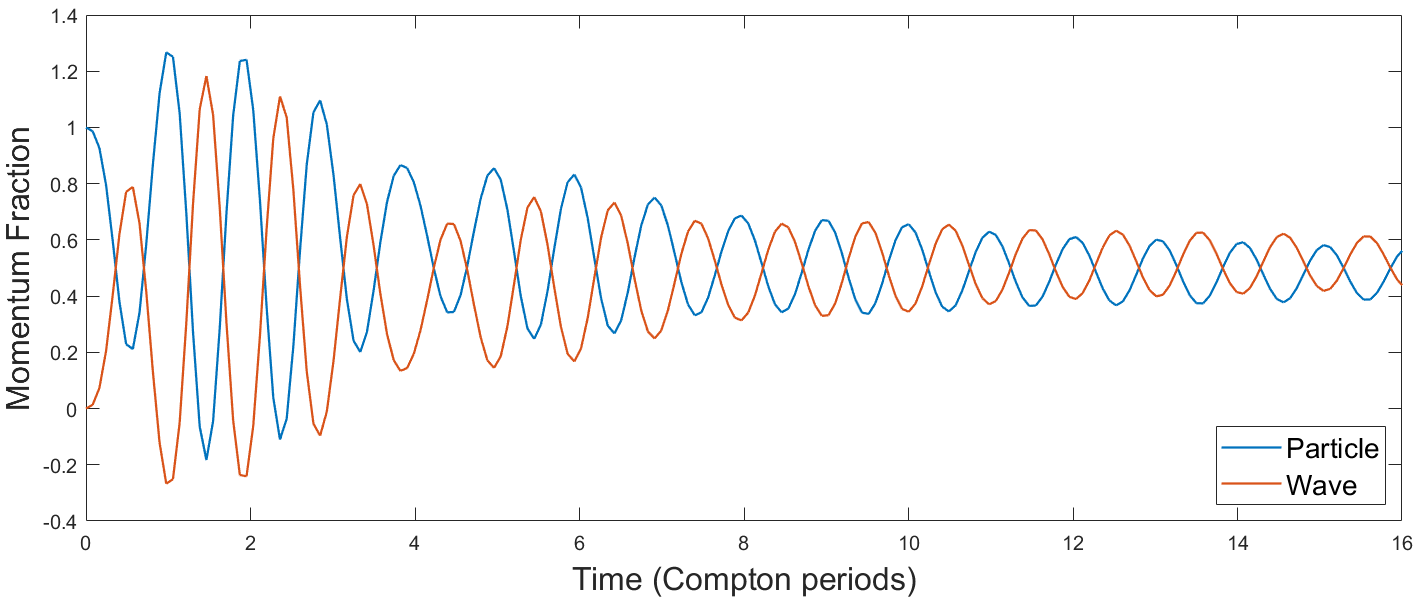}
	\caption{Exchange of horizontal (i.e., in-line) momentum between the particle and field after the particle is accelerated to an initial velocity $u_0 = 0.35$. The coupling constant is $b = 80.0$, corresponding to the lowermost curve in Figure \ref{fig:radiation}a. Note that the ``wave'' component of momentum incorporates both the local wavepacket of Section~\ref{sec:steadystate}, which gives rise to the virtual mass $\delta m$, and the radiating component detailed in Section~\ref{sec:radiation}. The momentum exchange is most pronounced for $t\lesssim 8$ Compton oscillations, after which 
    the system approaches a quasi-steady, periodic state. The former regime reflects the establishment of the robust wavepacket (\ref{eq:steadystategeneral}), and the latter the \emph{Zitterbewegung} detailed in Section~\ref{sec:oscillation}.}
 \label{fig:radiation2}
\end{figure}

Figure \ref{fig:radiation2} shows the exchange of (horizontal) momentum between particle and wave for the initial time period in Figure \ref{fig:radiation}a (for $b = 80.0$). Here, we see that the magnitude of oscillations dies down significantly after $t\sim 8$ Compton periods---i.e., as the particle approaches its steady-state velocity. After this point, the particle velocity continues oscillating at a lower amplitude, corresponding to the in-line \emph{Zitterbewegung} of Section~\ref{sec:oscillation}.

\vspace{-6pt}

Figure \ref{fig:radiation}b shows the quantity $|P_\text{part.}|/|P(t=0)|$ over a range of initial velocities $u_0$. This measures the fraction of momentum retained by the particle, which we use as a rough measure of the fraction of momentum \emph{transferred} to the wavepacket. The limitation of this metric is that momentum can \emph{either} be transferred to the wavepacket \emph{or} radiated away. Here, we see both effects. First, since the momentum fraction carried by the steady-state wavepacket is velocity-independent---as quantified in (\ref{eq:momentumfraction})---each curve in this figure is bounded above by $1 - |P_\text{wav}|/|P(t=0)|$. To continue, we assume that at the maximum point in each curve in Figure \ref{fig:radiation}b, the particle does not radiate \emph{any} momentum away. That is, for each $b$,
\[P_\text{wav} = P_\text{field}\]
at the maximum point on each curve. With this approximation, the best fit to $|P_\text{wav}|/|P_\text{total}|$ is given by
\begin{equation}\label{eq:virtmass}
    \delta m/m = \left(\frac{b}{163.8}\right)^2,
\end{equation}
with a maximum relative error of $0.5\%$. This represents a very close fit to our prediction $\delta m\propto b^2$.

Finally, looking beyond the maximum point of each curve, note that a significant amount of momentum is lost beyond the $(\delta m)\gamma u_s$ transferred to the wavepacket. As $b$ increases, the wavepacket itself grows, and more momentum is taken up by the virtual mass. Conversely, as $p_0$ decreases, more is radiated away on top of the wavepacket. We can understand this through Corollary \ref{cor:exchange}; the momentum transfer between particle and field is
\[\partial_\mu [T_\text{field}]^\mu_k = -\delta^3(\vec{q}-\vec{q}_p)\gamma^{-1}mb\partial_k\phi\propto \gamma^{-1}b^2,\]
so as $\gamma^{-1}b^2\to 0$ with increasing particle velocity or decreasing coupling constant, less momentum is available to radiate.

As a point of note, we see that the curve $\gamma^{-1}$ roughly cuts the system into two states: when accelerated (from rest) above a critical momentum $p^* = p^*(b)$, the particle settles quickly into a steady state with virtual momentum fraction
\[p_\text{virtual}/p_\text{tot}\sim \gamma^{-1}_{p^*} = (1+(p^*)^2)^{-1/2}.\]
\textls[-25]{This ``low-radiation regime'' characterizes the experiments above the black curve in Figure \ref{fig:radiation}b.} When accelerated below this critical momentum, however, the particle initially loses a substantial fraction of its momentum to radiation.

\subsection{Heisenberg Uncertainty, and the \texorpdfstring{$\lambda_c$}{lambda-c} Particle Cloud}\label{sec:uncertainty}

\begin{figure}[H]
	\begin{adjustwidth}{-\extralength}{0cm}
		\centering
		\begin{subfigure}{.5\linewidth}
			\centering
			\includegraphics[scale=0.4]{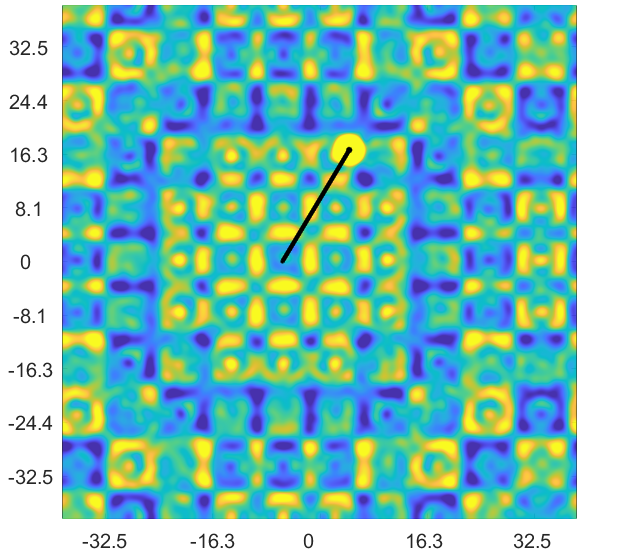}
			\caption{\centering}\label{fig:zitter3d_A}
		\end{subfigure}%
		\begin{subfigure}{.5\linewidth}
			\centering
			\includegraphics[scale=0.4]{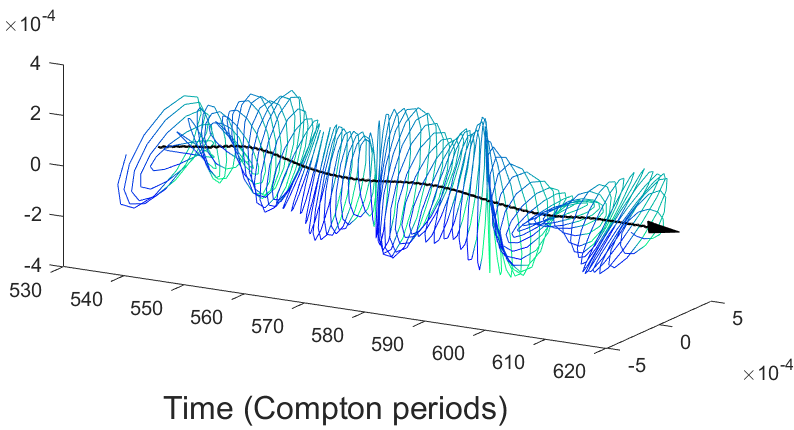}
			\caption{\centering}\label{fig:zitter3d_B}
		\end{subfigure}
	\end{adjustwidth}
	\caption{(\textbf{a}) A trajectory of the same form as Figure \ref{fig:radiation_ex}, but with $b = 66.7$ and initial velocity $\vec{u} = (0.15c,0.25c)$. By breaking the up-down symmetry of previous simulations, we can see particle vibrations in both in-line and transverse directions. Both axes are given in units of $\lambda_c$. (\textbf{b}) Particle vibrations about the mean trajectory in the $x$ and $y$ directions, scaled by $\lambda_c$. To resolve this motion, we performed a high-pass filter on the full trajectory. These vibrations form a moving particle cloud, which satisfies the Heisenberg uncertainty relation (\ref{eq:uncertainty}). Created with the MATLAB package \cite{MATLABarrow2023}. }\label{fig:zitter3d}
\end{figure}

\textls[-15]{Recall from Section~\ref{sec:radiation} that, in the case of a free particle, an in-line Zitter is excited by the phase waves washing over the particle. Now suppose the particle is not free, but confined to a finite geometry---we assume only that its waves wash over it from all directions, either reflected off of walls or generated by its periodic images. The resulting wave interference characterizes our periodic system in the long-time limit, as in Figure \ref{fig:radiation_ex}c,} but it is also expected to arise when the particle interacts with a variety of common quantum apparatuses (e.g., slits, corrals, or interferometers) or indeed during any position measurement.
In such confined geometries, Zitter occurs in \emph{all directions}, owing to the complex geometry of the incoming waves. The resulting motion is characterized by a region of scale $\lambda_c$ around the mean trajectory, about which the particle vibrates at a characteristic frequency $\gamma\omega_c$. We call this region the \emph{particle cloud}, 
whose form is shown in Figure \ref{fig:zitter3d}b.

\textls[-55]{We investigate this vibration in 1D by returning to the numerical experiments of Figure \ref{fig:spectrogram}, and focusing on their long-time limit,
wherein} previously-radiated waves are incoming from all directions. Amplitudes of these oscillations are given in Figure \ref{fig:oscillations} over a range of velocities and coupling constants. Namely, we calculate each amplitude as the $L^1$ norm (over frequency space) of the short-time Fourier transform discussed in Figure \ref{fig:spectrogram}, averaged over the last 1000 time-steps. Note that higher $L^p$ norms are bounded by this value up to a constant multiple, as the frequency spectra have (approximately) uniformly \linebreak  compact support.

In Figure \ref{fig:oscillations}b, we show best-fit curves of the form $n_b^{-1}\gamma^{-e_b}\lambda_c$ for each value of $b$. Here, $n_b$ and $e_b$ both decrease with $b$ for all values within the tested range. Critically, note that $e_b < 2$ for sufficiently large $b$. This gives an oscillation amplitude
\begin{equation*}
	\delta x \sim m^{-1}n_b^{-1}\gamma^{-e_b}
\end{equation*}
in any direction, where $e_b$ and $n_b$ are as in Figure \ref{fig:oscillations}. Using the characteristic oscillation frequency $\gamma\omega_c$, based on our discussion in Section~\ref{sec:radiation}, we find
\begin{equation}\label{eq:dv}
	\delta v\sim \gamma\omega_c\delta x\sim \gamma^{1-e_b}n_b^{-1}
\end{equation}
To estimate the corresponding momentum amplitude $\delta p$, suppose that the particle oscillates from a minimum momentum $p_0=m_\text{eff}\gamma_{0}v_0$ to a maximum $p_1=m_\text{eff}\gamma_{1}v_1$ in the direction of $\delta x$. Here, $m_\text{eff}$ is the \emph{effective mass} of both the particle and its steady-state wavepacket, as discussed in Section~\ref{sec:virtual}. Furthermore, note that $\gamma_0\geq \gamma_{v_0} :=(1-v_0^2)^{-1/2}$ and similarly $\gamma_1\geq \gamma_{v_1}$, with equality only if the particle is confined to move in one direction. In either case, the equality $dp = \gamma^3dv$ holds, where $\gamma$ is the \emph{total} Lorentz factor of the particle. Then, supposing transverse oscillations are small, we have
\[\delta p = p_1 - p_0 = m'\int_{v_0}^{v_1}dv\;\gamma^3\geq m_\text{eff}(v_1-v_0)\gamma^3_{\ovl{v}} = m_\text{eff}\gamma^3_{\ovl{v}}\delta v\]
from Jensen's inequality, using $\ovl{v} = (v_0+v_1)/2 + (\text{orthogonal})$ to denote the particle's mean velocity. In turn, our estimate (\ref{eq:dv}) gives
\[\delta p\sim m_\text{eff}\gamma^{4-e_b}n_b^{-1}.\]
\textls[-15]{Modeling $x$ and $p$ as monochromatic oscillations, we have $\sigma_x\sim \delta x/\sqrt{2}$ and $\sigma_p\sim\delta p/\sqrt{2}$, or}
\begin{empheq}[box=\fbox]{equation}\label{eq:uncertainty}
	\sigma_x\sigma_p\geq\tfrac{1}{2}(m_\text{eff}/m)\gamma^{4-2e_b}n_b^{-2}.
\end{empheq}
Finally, applying $e_b< 2$ gives \[\sigma_x\sigma_p\geq\frac{1}{2}(m_\text{eff}/m)n_b^{-2},\] 
or in dimensional form,
\[\sigma_x\sigma_p\geq\frac{1}{2}\hbar(m_\text{eff}/m) n_b^{-2}.\]
Given (\ref{eq:virtmass}), this inequality reduces to the traditional uncertainty principle for sufficiently large $b$. Note that a linear regression suggests that we should find $e_b<1$ at $b\sim 105$; in this stronger coupling regime, these oscillations would obey the stronger \emph{relativistic uncertainty principle} of Putra and Alrizal~\cite{PUTRA20191529}: $\sigma_x\sigma_p\geq\frac{1}{2}\hbar\gamma^2$.

We emphasize that the uncertainty relation (\ref{eq:uncertainty}) is somewhat different in nature to its counterpart in quantum mechanics.
Instead of an underlying property of a wave-like system, as in quantum theory, our uncertainty relation characterizes a \emph{classical} uncertainty of the Compton-scale particle dynamics, brought about by its waves interacting with a wall-bounded geometry. Averaging over the Compton timescale of the particle, the particle appears to take up a Compton-scale volume in phase space, reminiscent of quantum mechanics. However, in quantum mechanics, these scales can be squeezed in either position or momentum coordinates, giving rise to (a) highly localized states of indefinite momentum and (b) delocalised states of definite momentum. This type of squeezing does not appear to have an analogue in our uncertainty relation.

\section{Discussion}\label{sec:discussion}

The physical picture furnished by pilot-wave hydrodynamics has motivated 
a revisitation of de Broglie's mechanics. With a view to extending his 
double-solution theory~\cite{deBroglie1930,deBroglie1987}, we have 
presented a general Lagrangian framework for relativistic pilot-wave 
dynamics, and studied a particular limit of this framework in depth. 
Our framework has several structural advantages over the {\emph{HQFT} program} of Dagan, Durey, and Bush~\cite{Dagan2020,Durey2020b,Dagan2022}. First, our Lagrangian framework is Lorentz-covariant, and so in accord with the principles of special relativity. Second, it gives rise to a second-order equation for the particle trajectory, allowing classical dynamics to naturally emerge from our model in the $\hbar \rightarrow 0$ limit.

The dynamics of our system recovers a variety of behaviors familiar from de Broglie's theory, but with several key differences and advantages.
In both de Broglie's double-solution program and in HQFT~\cite{Dagan2020,Durey2020b,Dagan2022}, the internal particle vibration takes the form of an enforced particle clock at the Compton frequency. In our model, the particle clock emerges naturally from a time-invariant dynamical system, in the form of regular particle vibrations, or \emph{Zitterbewegung}. These vibrations take the form of in-line oscillations for a free particle, but may occur in all directions when the guiding wave is altered by bounding walls.

The Zitter of de Broglie's theory is subject to his \emph{harmony of phases}, according to which the internal particle oscillation is locked in phase with the oscillation of the pilot wave. In de Broglie's work, this phase-locking gave rise to the key relation $p = \hbar k$, the identification of the de Broglie wavelength $\lambda_{dB} = \hbar/(m\gamma u)$, and the adjustment of the particle oscillation to the Doppler-shifted frequency $\gamma^{-1}\omega_c$. The de Broglie relation also emerges from HQFT; however, there 
the particle momentum $\langle p^2\rangle = \hbar^2k^2$ is fixed by the wave-particle coupling constant. 
Our system recovers a \emph{dynamical} version of de Broglie's harmony of phases: not only does the pilot wave radiate at the de Broglie wavelength and the particle oscillate at the Doppler-shifted Compton frequency, but both effects update \emph{dynamically} as the particle accelerates, preserving the de Broglie relation $p=\hbar k$ even when the momentum $p$ changes.

The distinction between the wave form of the free particle deduced here and that of HQFT is noteworthy. While both are characterized by the particle moving in conjunction with a pilot wave with wavelength $\lambda_{dB}$, the manner in which the Compton wavelength appears is
markedly different.
In HQFT, waves of the Compton wavelength are continuously generated along the particle path. In our system, the Compton wavelength appears only in setting the scale of the Yukawa potential adjoining the particle. Thus, the pilot-wave form of the free particle is effectively monochromatic in the vicinity of the particle, as is the case in pilot-wave hydrodynamics. Nevertheless, wave interference can give rise to particle oscillations with a characteristic scale prescribed by the Compton wavelength.

Furthermore, the de Broglie waves in our system are not generated continuously by the particle, as they are in 
HQFT, but arise only as a result of particle acceleration. When the particle accelerates from one velocity $\vec{u}_0$ to another velocity $\vec{u}_1$, it spawns a new, continuous source of waves, which continues at a velocity $ \vec{u}_{source}\sim\vec{u}_0$ along the \emph{extrapolated, original trajectory} of the particle. We note that a comparable physical picture was explored by Fort~\cite{Fort_2013} in his investigation of orbiting `inertial walkers', a theoretical abstraction of the walking-droplet system in which wave sources do not strictly follow the particle, but instead extend in rays along the instantaneous particle paths. In the first simple example considered here, when the particle accelerates 
from rest, waves are generated continuously \emph{from the point of acceleration}. After traveling an appreciable distance, the free particle thus surfs on an approximately planar wavefront of the de Broglie wavelength at nearly constant speed, as its in-line Zitter diminishes with time. 

Another important feature of the present system is that our particle always travels with a Yukawa wavepacket, which stores energy in a region of radius $\sim\lambda_c$ adjoining the particle. We have shown that this energy takes the form of a wave-induced \emph{virtual mass} that augments the particle's inertial mass to $m\mapsto m + \delta m$. A similar wave-induced mass has been reported and rationalized in pilot-wave hydrodynamics and accounts for the anomalously large radii of inertial orbits in that system~\cite{bush_boost_2014}. A critical difference is that in the hydrodynamic system, the wave-induced mass is \emph{velocity-dependent}, while ours is not. This difference simply reflects the Lorentz-covariance of the present model: the virtual mass of our system, as a Lorentz scalar, \emph{cannot} depend upon the particle velocity.


Finally, by combining the virtual mass and the multi-directional Zitterbewegung of our system, we have recovered a dynamical version of the Heisenberg uncertainty principle: \[\sigma_x\sigma_p \geq K\hbar/2,\]
where $K$ depends only on the coupling constant $b$,
and $\sigma_x$ and $\sigma_p$ are standard deviations of the position and momentum in any one direction. For sufficiently large particle-field coupling constants, $b$, this inequality reduces to the standard uncertainty principle of quantum mechanics. 

While we have focused here on a particular limit of our Lagrangian framework, our model allows for the consideration of other wave-particle couplings that will be explored elsewhere. In future studies, we shall adopt the present model to perform simulations of the classic single- and double-slit diffraction experiments, and compare the emergent statistics with those predicted by standard quantum mechanics.
We shall also examine the case in which the particle moves in response to the gradient of the wave's phase, and demonstrate that this particular variant allows for a recovery of Bohmian mechanics and a compelling model of wavefunction collapse following a position measurement.

{Finally, it is perhaps worth reminding the reader that the mathematical framework developed here is not an attempt to replace modern quantum theory; moreover, we do not claim that it will reproduce the results of quantum field theory. However, it does represent a rigorous, modern version of the mechanics envisaged by de Broglie in his double-solution program. As such, we hope that it may prove fruitful in exploring the boundary between classical and quantum behavior.}


\vspace{6pt} 



\authorcontributions{
Conceptualization, D.D. and J.W.M.B.; methodology, D.D.; software, D.D.; validation, D.D.; formal analysis, D.D.; investigation, D.D.; resources, J.W.M.B.; data curation, D.D.; \linebreak  writing---original draft preparation, D.D.; writing---review and editing, D.D. and J.W.M.B.;\linebreak visualization, D.D.; supervision, J.W.M.B.; project administration, J.W.M.B.; funding acquisition, J.W.M.B. All authors have read and agreed to the published version of the manuscript.}

\funding{\textls[-15]{This research was funded by the National Science Foundation through grant CMMI-2154151.}} 

\dataavailability{All data for the numerical experiments of the present work is available in our GitHub repository, at \url{https://github.com/ddarrow90/hqftii} (accessed on 18 December 2023).}

\acknowledgments{The first author would like to thank MathWorks, whose generous support through the MathWorks Fellowship has made this research possible.}

\conflictsofinterest{The authors declare no conflict of interest.}

\appendixtitles{yes} 
\appendixstart
\appendix

\section[\appendixname~\thesection]{Derivation of the Variational Theory}\label{sec:variationproofs}

\subsection[\appendixname~\thesubsection]{Variational Proofs in the General Case}\label{sec:variationproofs_A}
We here present the proofs of our variational results, i.e., the pilot-wave version of the Euler--Lagrange equations and of Noether's theorem, respectively. See Lemmas \ref{lem:eullag} and \ref{lem:noether} for the statement of these results.

\begin{proof}[Proof of Lemma \ref{lem:eullag}]
	\textls[-35]{By linearity of the functional derivative, the action is extremized over variations in both $\phi$ and $q_p$ when it is extremized over variations in each individually. As $\delta_{q_p}\mathcal{S}_\text{field} = 0$, the particle's trajectory equation follows from the classical Euler--Lagrange equations.}
	
	For the field equation, we make use of the Euler--Lagrange equations for a scalar field:
	\[\delta_\phi\mathcal{L} = \partial_\mu\delta_{\phi_\mu}\mathcal{L}.\]
	Specifically, viewing the particle and interaction Lagrangians as singular Lagrangian fields over spacetime, we apply the above to find
	\begin{align*}
		\delta_\phi(\mathcal{L}_\text{field} + \delta^3(q-q_p)\mathcal{L}_\text{part.})= \partial_\mu\delta_{\phi_\mu}(\mathcal{L}_\text{field} + \delta^3(q-q_p)\mathcal{L}_\text{part.}).
	\end{align*}
	On the left-hand side, the variational derivative commutes with the delta function, giving two terms of the field equation. On the right-hand side, the variational derivative commutes similarly. Combining with the fact that
	\[\partial_t\delta^3(q-q_p) = -\vec{u}\cdot\nabla\delta^3(q-q_p)\]
	gives the desired field equation.
\end{proof}

\begin{proof}[Proof of Lemma \ref{lem:noether}]
	Let $\mathcal{L}'_\text{field} = \mathcal{L}_\text{field} + \eps\partial_\mu\Lambda^\mu_\text{field}$ and $\mathcal{L}'_\text{part.} = \mathcal{L}_\text{part.} + \eps d_t\Lambda_\text{part.}$. We deal with field and singular components separately, allowing us to quantify the exchange of conserved quantities between them. Beginning with the field Lagrangian, we find (to first order in $\eps$) that
  \begin{adjustwidth}{-\extralength}{0cm}
    \vspace{-\baselineskip}
	\begin{align*}
		\eps^{-1}(\mathcal{L}_\text{field}' - \mathcal{L}_\text{field}) &= (\delta_{\phi}\mathcal{L}_\text{field})\Psi + (\delta_{\phi_\mu}\mathcal{L}_\text{field})\partial_\mu\Psi\\
		&= (\delta_{\phi}\mathcal{L}_\text{field} - \partial_\mu\delta_{\phi_\mu}\mathcal{L}_\text{field})\Psi + \partial_\mu((\delta_{\phi_\mu}\mathcal{L}_\text{field})\Psi)\\
		&= \Psi\delta^3(\vec{q}-\vec{q}_p)(\partial_\mu\delta_{\phi_\mu}\mathcal{L}_\text{part.} - \delta_\phi\mathcal{L}_\text{part.}) + \partial_\mu((\delta_{\phi_\mu}\mathcal{L}_\text{field})\Psi)- \Psi\nabla\delta^3(\vec{q}-\vec{q}_p)\cdot (\vec{u}\delta_{\phi_0} -\delta_{\nabla\phi})(\mathcal{L}_\text{part.}),
	\end{align*}
 \end{adjustwidth}
	giving
  \begin{adjustwidth}{-\extralength}{0cm}
  \vspace{-\baselineskip}
	\begin{align}\label{eq:cons1}
		\begin{split}
			\partial_\mu(\Lambda_\text{field}^\mu -(\delta_{\phi_\mu}\mathcal{L}_\text{field})\Psi) = \Psi\delta^3(\vec{q}-\vec{q}_p)(\partial_\mu\delta_{\phi_\mu}\mathcal{L}_\text{part.} - \delta_\phi\mathcal{L}_\text{part.})- \Psi\nabla\delta^3(\vec{q}-\vec{q}_p)\cdot (\vec{u}\delta_{\phi_0} -\delta_{\nabla\phi})(\mathcal{L}_\text{part.}).
		\end{split}
	\end{align}
  \end{adjustwidth}
	For the singular component, we similarly find
  \begin{adjustwidth}{-\extralength}{0cm}
  \vspace{-\baselineskip}
	\begin{align*}
		\eps^{-1}(\mathcal{L}'_\text{sing}-\mathcal{L}_\text{sing}) &= (\delta_\phi\mathcal{L}_\text{part.})\Psi + (\delta_{\phi_\mu}\mathcal{L}_\text{part.})\partial_\mu\Psi + (\delta_{q_p^k}\mathcal{L}_\text{part.})Q^k + (\delta_{u^k}\mathcal{L}_\text{part.})d_tQ^k\\
		&= (\delta_\phi\mathcal{L}_\text{part.}-\partial_\mu\delta_{\phi_\mu}\mathcal{L}_\text{part.})\Psi + \partial_\mu((\delta_{\phi_\mu}\mathcal{L}_\text{part.})\Psi)+ (\delta_{q_p^k}\mathcal{L}_\text{part.}-d_t\delta_{u^k}\mathcal{L}_\text{part.})Q^k + d_t((\delta_{u^k}\mathcal{L}_\text{part.})Q^k).
	\end{align*}
  \end{adjustwidth}
	Inserting the generalized Euler--Lagrange Equation (\ref{eq:noncons}) as well, we find
  \begin{adjustwidth}{-\extralength}{0cm}
    \vspace{-\baselineskip}
	\begin{align*}d_t(\Lambda_\text{part.} - (\delta_{u^k}\mathcal{L}_\text{part.})Q^k) = \partial_\mu((\delta_{\phi_\mu}\mathcal{L}_\text{part.})\Psi) + (\delta_\phi\mathcal{L}_\text{part.}-\partial_\mu\delta_{\phi_\mu}\mathcal{L}_\text{part.})\Psi -F_kQ^k.
 \end{align*}
  \end{adjustwidth}
	Before combining the particle and field components, we note that
  \begin{adjustwidth}{-\extralength}{0cm}
  \vspace{-\baselineskip}
	\begin{align*}
		\delta^3(\vec{q}-\vec{q}_p)\partial_\mu((\delta_{\phi_\mu}\mathcal{L}_\text{part.})\Psi) &= \partial_\mu\left(\delta^3(\vec{q}-\vec{q}_p)(\delta_{\phi_\mu}\mathcal{L}_\text{part.})\Psi\right)+ \Psi\nabla\delta^3(\vec{q}-\vec{q}_p)\cdot(\vec{u}\delta_{\phi_0} -\delta_{\nabla\phi})(\mathcal{L}_\text{part.}).
	\end{align*}
  \end{adjustwidth}
	With this in mind, we combine the above with (\ref{eq:cons1}) to recover
	\[
	\partial_\mu(\Lambda_\text{field}^\mu -(\delta_{\phi_\mu}\mathcal{L}_\text{total})\Psi) +\delta^3(\vec{q}-\vec{q}_p)d_t(\Lambda_\text{part.} - (\delta_{u^k}\mathcal{L}_\text{part.})Q^k)= -\delta^3(\vec{q}-\vec{q}_p)F_kQ^k.
	\]
	We note that $d_t\delta^3(\vec{q}-\vec{q}_p) = (\vec{u}\cdot\nabla - \vec{u}\cdot\nabla)\delta^3(\vec{q}-\vec{q}_p) \equiv 0$, so we can commute the time derivative with the delta function. Furthermore, we can substitute $d_t = u^\mu\partial_\mu$, as the latter is the material derivative along the particle trajectory. Note that this substitution works whether we interpret $Q^k$ as a function of $q_p = q_p(t)$ \emph{or} as a function of the spatial coordinates $q$, as we can see by the chain rule. The theorem follows.
\end{proof}

\subsection[\appendixname~\thesubsection]{Stress-Energy Conservation in the AM System}\label{sec:variationproofs_B}
We now restrict our attention to the \emph{amplitude-modulated} (AM) system introduced in Section~\ref{sec:amplitude}, and derive the balance law (\ref{eq:SEcontinuity}). In short, this approach guarantees an exact conservation of momentum and a \emph{near} conservation of the total energy budget. The latter evolves over time in a manner proportional to $d_t\phi(\vec{q}_p) = u^\mu\partial_\mu\phi(\vec{q}_p)$.

We consider the specific pilot-wave framework introduced in Appendix \ref{sec:particle_equation_B}. That is, our Lagrangians take the following forms:
\begin{align}\label{eq:lagrangians}
	\begin{split}
		\mathcal{L}_\text{field} &= \tfrac{1}{2}\left(\partial^\mu\phi\partial_\mu\phi - m^2\phi^2\right),\\
		\mathcal{L}_\text{part.} &= -\gamma^{-1} m - b\tau u^\mu\partial_\mu\phi,
	\end{split}
\end{align}
\textls[-15]{where $b$ is a fixed coupling constant and $\tau$ is the proper time along the particle's trajectory (synchronized at $t=0$, say). Recall that, while the above Lagrangians do not obey the general form (\ref{eq:oureqs}), they give rise to the AM system when we further apply the non-conservative force}
\[\vec{F} = - mbd_t\left((\delta_{\vec{u}}\tau)u^\mu\partial_\mu\phi(q_t)\right),\]
as in the equations (\ref{eq:noncons}). 

Now, suppose we have a constant spatial translation $q^k\mapsto q^k + \eps Q^k$, so that 
\[(q_p)^k\mapsto (q_p')^k:=(q_p)^k + \eps Q^k,\qquad \phi\mapsto \phi':=\phi - \eps Q^k\partial_k\phi.\]
With this spatial translation, we find that $\mathcal{L}_\text{part.}$ remains invariant. Indeed, 
\[\mathcal{L}'_\text{part.}=-\gamma^{-1}m - b\tau u^\mu\partial_\mu\phi'(q_p'),\]
but $\phi'(q_p') = \phi(q_p)$ and, since the velocity $u$ is invariant, neither $\gamma$ nor $\tau$ changes. The field Lagrangian is translated as
\[\mathcal{L}'_\text{field} = \mathcal{L}_\text{field} - \eps Q^k\partial_k\mathcal{L}_\text{field}=\mathcal{L}_\text{field} - \eps Q^k\partial_\mu \delta^\mu_k\mathcal{L}_\text{field},\]
as it has no explicit spatial dependence. Applying Lemma \ref{lem:noether} above, we recover
\begin{adjustwidth}{-\extralength}{0cm}
\vspace{-\baselineskip}
\begin{align*}
	T_k^{\mu} &= -\tfrac{1}{2}\delta^\mu_k(\partial_\alpha\phi\partial^\alpha\phi - m^2\phi^2) + (\partial^\mu\phi - \delta^3(\vec{q}-\vec{q}_p)mb\tau u^\mu)\partial_k\phi + u^\mu\delta^3(\vec{q}-\vec{q}_p)(m\gamma u_k + mb\tau\partial_k\phi - J_k )\\
	&= -\tfrac{1}{2}\delta^\mu_k(\partial_\alpha\phi\partial^\alpha\phi - m^2\phi^2) + \partial^\mu\phi\partial_k\phi+ u^\mu\delta^3(\vec{q}-\vec{q}_p)(m\gamma u_k- J_k ),
\end{align*}
\end{adjustwidth}
where $J_k$ is given by (\ref{eq:impulse}).
Supposing that our particle evolves under the non-conservative equations (\ref{eq:noncons}) with $\vec{F}=d_t\vec{J}= u^\mu\partial_\mu \vec{J}$, as we do in Appendix~\ref{sec:particle_equation_B}, we find that the $J_k$ term disappears in the balance equation: that is,
\[\tilde{T}_k^\mu = -\tfrac{1}{2}\delta^\mu_k(\partial_\alpha\phi\partial^\alpha\phi - m^2\phi^2) + \partial^\mu\phi\partial_k\phi+ u^\mu\delta^3(\vec{q}-\vec{q}_p)m\gamma u_k\]
satisfies $\partial_\mu\tilde{T}_k^\mu = 0$ exactly.

Moving on to time translations, suppose that $t\mapsto t + \eps Q^0$ and spatial coordinates are left fixed. Then we have instead that
\[(q'_p)^k = (q_p)^k - \eps Q^0u^k,\qquad \phi' = \phi - \eps Q^0\partial_0\phi.\]
The field Lagrangian translates as before:
\[\mathcal{L}'_\text{field} = \mathcal{L}_\text{field} - \eps Q^0\partial_0\mathcal{L}_\text{field},\]
although the singular Lagrangian does not. If $\mathcal{L}_\text{part.}$ had no explicit time dependence, we would find
\[\mathcal{L}_\text{part.}' \stackrel{?}{=} \mathcal{L}_\text{part.} - \eps Q^0d_t\mathcal{L}_\text{part.}.\]
However, by replacing the passive coordinate transformation with an active transformation of $q_p$ and $\phi$, we see that we must remove a term corresponding to $\partial_t\tau=\gamma^{-1}$ from \linebreak  the above:
\[\mathcal{L}_\text{part.}' = \mathcal{L}_\text{part.} - \eps Q^0d_t\mathcal{L}_\text{part.} - \eps Q^0mb\gamma^{-1} u^\mu\partial_\mu\phi.\]
Write $R(t) = \int_0^tdt'\;\gamma^{-1} u^\mu\partial_\mu\phi(q_p)$, so that
\[\mathcal{L}_\text{part.}' = \mathcal{L}_\text{part.} - \eps Q^0d_t\left(\mathcal{L}_\text{part.} + mbR(t)\right).\]
Then we find
\begin{adjustwidth}{-\extralength}{0cm}
\vspace{-\baselineskip}
\begin{align*}
	T^\mu_0 &= -\tfrac{1}{2}\delta^\mu_0(\partial_\alpha\phi\partial^\alpha\phi - m^2\phi^2) + \partial_0\phi\partial^\mu\phi - \delta^3(\vec{q}-\vec{q}_p)mb\tau u^\mu\partial_0\phi\\
	&\qquad 
	+ u^\mu\delta^3(\vec{q}-\vec{q}_p)(m\gamma^{-1} + mb\tau u^\alpha\partial_\alpha\phi + mbR(t)
	- u^km\gamma u_k - u^kmb\tau\partial_k\phi(q_p) + u^kJ_k)\\
	&= -\tfrac{1}{2}\delta^\mu_0(\partial_\alpha\phi\partial^\alpha\phi - m^2\phi^2) + \partial_0\phi\partial^\mu\phi - \delta^3(\vec{q}-\vec{q}_p)mb\tau u^\mu\partial_0\phi \\
	&\qquad+ u^\mu\delta^3(\vec{q}-\vec{q}_p)(m(\gamma^{-2}-u^ku_k)\gamma + mb\tau \partial_0\phi + mbR(t) + u^kJ_k)\\
	&= -\tfrac{1}{2}\delta^\mu_0(\partial_\alpha\phi\partial^\alpha\phi - m^2\phi^2) + \partial_0\phi\partial^\mu\phi+ u^\mu\delta^3(\vec{q}-\vec{q}_p)(m\gamma u_0 + mbR(t) + u^kJ_k).
\end{align*}
\end{adjustwidth}
As before, the $J_k$ term cancels in the balance equation, so the modified vector
\[\tilde{T}^\mu_0 = -\tfrac{1}{2}\delta^\mu_0(\partial_\alpha\phi\partial^\alpha\phi - m^2\phi^2) + \partial_0\phi\partial^\mu\phi+ u^\mu\delta^3(\vec{q}-\vec{q}_p)m\gamma u_0\]
satisfies
\[\partial_\mu\tilde{T}^\mu_0 = -\partial_\mu(u^\mu\delta^3(\vec{q}-\vec{q}_p)mbR(t))=-mb\delta^3(\vec{q}-\vec{q}_p)\gamma^{-1} u^\mu\partial_\mu\phi(q_p),\]
where we differentiated $u^\mu\partial_\mu R = d_t R$. We have thus deduced the desired energy balance relation.

\section[\appendixname~\thesection]{Deriving the Approximate Equation (\ref{eq:equations_am2})}\label{sec:particle_equation}
We here provide two different derivations of the \emph{amplitude-modulated} (AM) system of Section~\ref{sec:amplitude}. The first employs a scaling argument to demonstrate that the AM system approximately fits the form (\ref{eq:oureqs}) demanded by our general Lagrangian framework. The second deviates slightly from this framework, employing a non-conservative variational approach to reach the AM system. The latter derivation has two key advantages. First, it gives rise to useful stress-energy (and thus angular momentum) conservation relations, derived in Section~\ref{sec:AMconservation}.
Second, it allows us to obtain bounds on the approximate Lorentz covariance of the AM system, derived in Appendix~\ref{sec:covariance}.
\subsection[\appendixname~\thesubsection]{An Approximate Route to the AM System}\label{sec:particle_equation_A}
Below, we use the notation $\langle A\rangle$ to denote a \emph{scale estimate} of a given quantity $A$, rather than a mean. If the estimates $\langle\gamma\rangle$ and $\langle u\rangle$ of the Lorentz factor and velocity, respectively, arise from the same mean state, this means that we can take $\langle\gamma u\rangle = \langle\gamma\rangle\langle u\rangle$ to hold exactly.

To derive the equation (\ref{eq:equations_am2}) for the AM dynamics, take $\phi:=\op{Re}\phi$ as before and
\[\sigma(\phi) = 1 + b^2/4\pi + b\phi,\]
so the full particle equation reads
\begin{equation*}
	d_t\left(m\gamma (1 + b^2/4\pi + b\ovl{\phi})\vec{u}\right) = bm\gamma^{-1}\nabla\ovl{\phi}.
\end{equation*}
Recall that $\ovl{\phi}$ denotes the \emph{continuous component} of $\phi$, as laid out in Section~\ref{sec:continuous}.
Splitting $\phi = \phi_\text{rad} + \phi_\text{wav}$ into radiative and wavepacket terms, where $\phi_\text{wav}$ is given by (\ref{eq:steadystategeneral}), we deduce the continuous component of $\phi$:
\[\ovl{\phi} = \ovl{\phi}_\text{rad} + \ovl{\phi}_\text{wav} = \frac{b}{4\pi m}\cdot\lim_{r\to 0}\frac{1}{4\pi}\int_{S^2} d\xi\;\partial_r e^{-mr} + \ovl{\phi}_\text{wav} = -\frac{b}{4\pi} + \phi_\text{wav},\]
\textls[-15]{by using the fact that $\phi_\text{wav}$ is continuous and that $\phi\mapsto\ovl{\phi}$ is Lorentz-invariant (from Section~\ref{sec:continuous})} on our domain. The latter allows us to carry out the above computation in the reference frame of the particle, without loss of generality. In turn, this gives the modified (exact!) trajectory equation
\begin{equation}\label{eq:crapequation}
	d_t\left(m\gamma (1 + b\phi_\text{rad})\vec{u}\right) = bm\gamma^{-1}\nabla\phi_\text{rad}.
\end{equation}
From Section~\ref{sec:oscillation}, we know that the field tends to vary at the length and time scales
\[L:=\ovl{\lambda}_\text{dB}=\frac{1}{m\langle\gamma u\rangle},\qquad T:= \frac{1}{m\langle\gamma\rangle}= L\langle u\rangle.\]
We use these scales to define the dimensionless derivatives $d_{t*} = Td_t$, $\partial_{t^*} = T\partial_t$, $\nabla^* = L\nabla$, and $\Delta^* = L^2\nabla^2$. To estimate $\phi_\text{rad}$, recall from Section~\ref{sec:oscillation} that momentum fluctuations of the particle are of the order
\[\delta p \sim \gamma^{4-e_b}n_b^{-1}mc,\]
\textls[-15]{where $e_b$ and $n_b$ are given in Figure \ref{fig:oscillations}. However, from the trajectory equation (\ref{eq:equations_am2}), we also know that}
\[\frac{\delta p}{mT}\sim \gamma^{-1}bL^{-1}\langle\phi_\text{rad}\rangle,\]
where $\langle\phi_\text{rad}\rangle$ is our desired $\phi_\text{rad}$ scale. 
Equating the two estimates yields
\begin{equation*}
	\langle\phi_\text{rad}\rangle=\frac{\langle\gamma^{5-e_b}\rangle }{\langle u\rangle bn_b},
\end{equation*}
which we plug into (\ref{eq:crapequation}) to find
\begin{equation*}
	d_t\left(m'\gamma \vec{u}\right) = bm\gamma^{-1}\nabla\ovl{\phi}, \qquad m' = m + O(m\gamma^{5-e_b}/n_bu).
\end{equation*}
In the range $13.3\leq b\leq 80.0$ used in our simulations, and assuming that $\langle u\rangle$ is neither too small nor too large, this corresponds to an error in the mass of at most a few percent. Dropping this oscillating mass in favor of its mean, we recover the amplitude-modulated system of Section~\ref{sec:amplitude}.



\subsection[\appendixname~\thesubsection]{An Exact, Non-Conservative Route to the AM System}\label{sec:particle_equation_B}
Although we show in Appendix \ref{sec:particle_equation} that the Equations (\ref{eq:equations_am1}) and (\ref{eq:equations_am2}) follow approximately from the action (\ref{eq:oureqs}), we can deduce exact conservation laws by deriving the AM model from an alternate, \emph{non-conservative} set of Euler--Lagrange equations.
In the language of Lemma \ref{lem:eullag}, we temporarily introduce the new Lagrangians
\begin{align*}
	\begin{split}
	\mathcal{L}_\text{field} &= \tfrac{1}{2} m^2\left(\partial^\mu\phi\partial_\mu\phi - m^2\phi^2\right),\\
	\mathcal{L}_\text{part.} &= -\gamma^{-1} m - mb \tau u^\mu\partial_\mu\phi,
	\end{split}
\end{align*}
where $\tau$ is the proper time along the particle trajectory. In the language of our general framework (\ref{eq:lagrangians}), these Lagrangians roughly identify the coupling $\sigma$ as
\[\sigma = \sigma(u^\mu,\partial_\mu\phi) = 1 + b\tau \gamma u^\mu\partial_\mu\phi.\]
As $\sigma$ depends on $u^\mu$ and $\partial_\mu\phi$, rather than just $\phi$, this coupling does \emph{not} fit our proposed framework. With that said, we can still identify dynamical equations and conservation laws from Lemmas \ref{lem:eullag} and \ref{lem:noether}.

We derive the field and particle equations in turn, as the latter (despite appearances from Lemma \ref{lem:eullag}) is more involved. In applying Lemma \ref{lem:eullag} to our particular system, note that
\[\delta_{\nabla\phi}\mathcal{L}_\text{part.} = -mb \tau\vec{u} = \vec{u}\delta_{\phi_0}\mathcal{L}_\text{part.},\]
so terms involving $\nabla\delta^3$ vanish from the field equation. We then evaluate
\[(\partial_\mu\delta_{\phi_\mu}-\delta_\phi)\mathcal{L}_\text{field} = m^2\partial_\mu\partial^\mu\phi + m^4\phi,\]
and
\begin{align*}
	(\partial_\mu\delta_{\phi_\mu}-\delta_\phi)\mathcal{L}_\text{part.} &= -\partial_\mu(mb\tau u^\mu)= -mb\partial_t \tau = -\gamma^{-1}mb,
\end{align*}
noting that $u^0\equiv 1$ and $d\tau = \gamma^{-1}dt$. Putting this together, we find the equation (\ref{eq:equations_am1}). Turning now to the particle trajectory equation, we find
\begin{equation}\label{eq:particlestepzero}
	\delta_{\vec{q}_p}\mathcal{L}_\text{part.} = -mb\tau u^\mu\partial_\mu\nabla\phi(q_t),
\end{equation}
\begin{equation}\label{eq:particlestepone}
	\delta_{\vec{u}}\mathcal{L}_\text{part.} =  m\gamma\vec{u}- mb\tau\nabla\phi(q_t) - mb(\delta_{\vec{u}}\tau)u^\mu\partial_\mu\phi(q_t).
\end{equation}
At this stage, we rewrite the final term above using
\[\delta_{u^j}\tau = \tpdrv{\gamma u^{\alpha}}{u^j}\delta_{\gamma u^\alpha}\tau=\gamma(\delta^\alpha_j - \gamma^2u^\alpha u_j)\delta_{\gamma u^\alpha}\tau,\]
which reveals it to be a Lorentz-invariant \emph{impulse of interaction}:
\begin{equation}\label{eq:impulse}
	J_j=-(\delta^\alpha_j - \gamma^2u^\alpha u_j)(\delta_{\gamma u^\alpha}\tau) mb\gamma u^\mu\partial_\mu\phi(q_t).
\end{equation}
Defining $\vec{F} = \frac{d}{dt}\vec{J}$ and differentiating (\ref{eq:particlestepone}) yields
\begin{align*}
	d_t\delta_{\vec{u}}\mathcal{L}_\text{part.}
	=d_t\left(m\gamma\vec{u}\right)- \gamma^{-1}mb\nabla\phi(q_t) + mb\tau u^\mu\partial_\mu\nabla\phi(q_t) + \vec{F}.
\end{align*}
Subtracting off (\ref{eq:particlestepzero}) and adding $\vec{F}$ as a non-conservative force, as in (\ref{eq:noncons}), we recover the trajectory equation (\ref{eq:equations_am2}).

Although we could have proceeded with a simpler Lagrangian, inspired by (\ref{eq:action}), there are two key benefits to our choice. First, the non-conservative component $\vec{F}$ vanishes in the non-relativistic limit of our system, reducing our equations again to exact Euler--Lagrange equations. Second, we will see that the remaining effect of $\vec{F}$ applies \emph{exclusively} to the energy budget of the system, leaving both momentum and angular momentum untouched. We derive these conservation laws in detail in Section~\ref{sec:AMconservation}.

\subsection{Approximate Lorentz Covariance}\label{sec:covariance} One caveat with the non-conservative derivation of Appendix~\ref{sec:particle_equation_B} is that the vector $F_\alpha:=\tfrac{d}{dt}J_\alpha$ does not take the correct form for a relativistically compatible force. While $\gamma F_\alpha$ is Lorentz-covariant, it does not satisfy the requirement $\gamma F_\alpha u^\alpha = 0$ for the particle to remain on-shell. The result is that (\ref{eq:equations_am2}) is not exactly Lorentz-covariant if $b\neq 0$, as it cannot be extended to an equation on four-vectors.

In this appendix, we quantify the error in this \emph{approximate} Lorentz-covariance, and determine where we can still make claims based on Lorentz transformations. Approximate covariance is used in three claims throughout our paper. First, in Section~\ref{sec:steadystate}, we apply a Lorentz transformation to the steady-state solution (\ref{eq:steadystate}) in order to deduce the form of the traveling wavepacket (\ref{eq:steadystategeneral}). Second, in Section~\ref{sec:oscillation}, we use this principle to argue that the Zitterbewegung frequency $\gamma^{-1}\omega_c$ is obtained after an acceleration from an \emph{arbitrary} initial walking state. Finally, in Section~\ref{sec:radiation}, we use this principle to derive the form of the accompanying waveform, as illustrated in Figure \ref{fig:diagram}.


In our laboratory frame, we have the trajectory equation
\[d_t\left(m\gamma\vec{u}\right) = \gamma^{-1} mb\nabla\phi(q_t).\]
First consider a mean velocity $\vec{u} = u\hat{e}_1$, and suppose we perform a Lorentz boost with velocity $\vec{v} = v\hat{e}_1$, resulting in the new four-momentum
\[\gamma' = \gamma_v\gamma(1-uv),\qquad \gamma'u' = \gamma_v\gamma(u-v).\]
On the one hand,
\[d_\tau (m\gamma'u') = \gamma_v(1-uv)d_\tau (m\gamma u) = (\gamma'/\gamma)d_\tau (m\gamma u),\]
or equivalently,
\[d_{t'}(m\gamma'u') = d_t(m\gamma u)\]
with the new time coordinate $t'$.
On the other hand, note that
\[\gamma^{-1} = (\gamma_v\gamma'(1+u'v))^{-1}\]
and
\[\partial_{x}\phi = \gamma_v(\partial_{x'} + v\partial_{t'})\phi.\]
Suppose $\phi$ has the local phase velocity $c^2\hat{e}_1/s$ in the boosted frame, corresponding to a de Broglie wave with group velocity $s\hat{e}_1$. In this setting, $s\partial_{t'}\phi(q_p) = \partial_{x'}\phi(q_p)$, and we recover a final equation
\[d_{t'}\left(m\gamma' u'\right) = (\gamma')^{-1} mb\frac{1-v/s}{1+u'v}\partial_{x'}\phi(q_t).\]
With this equation in hand, consider the case of a particle starting from rest (in the original, or laboratory, frame) and accelerating in the $\hat{e}_1$ direction. As we demonstrate in Section~\ref{sec:radiation}, this acceleration gives rise to quasi-monochromatic radiation of wavelength $\lambda_{dB}$ in a region around the particle, and thus $s = \langle u'\rangle$. Writing $u' = s + \eps(t')$, with $\eps$ encapsulating the small-amplitude Zitterbewegung of Section~\ref{sec:oscillation}, we find
\[d_{t'}\left(m\gamma' u'\right) = (\gamma')^{-1} m\tilde{b}(\partial_{x'}\phi(q_t))(1 + O(\eps v)),\]
where
\[\tilde{b} = \frac{1-v/s}{1+vs}b.\]
As shown in Figure \ref{fig:oscillations}, $|\eps|\sim n_b^{-1}\leq 1/21.0$ in all of our numerical experiments. Even for moderate-to-large initial velocities $v$, our trajectory thus satisfies the modified equation
\[d_{t'}\left(m\gamma' u'\right) = (\gamma')^{-1} m\tilde{b}\partial_{x'}\phi\]
in the boosted frame. Since the field equation (\ref{eq:equations_am1}) is invariant only \emph{without} a change in $b$, we now define the following rescalings of both $b$ and $\phi$:
\begin{equation}\label{eq:lorentzrescaling}
    b' = \sqrt{\tilde{b} b},\qquad \phi' = \sqrt{\tilde{b}/b}\phi.
\end{equation}
Under these rescalings, $\phi'$ satisfies the transformed equation
\[(\partial^\mu\partial_\mu + m^2)\phi' = (\gamma')^{-1}mb'\delta^3(\vec{q}-\vec{q}_p), \]
and $u'$ satisfies the transformed equation
\[d_{t'}\left(m\gamma' u'\right) = (\gamma')^{-1} mb'\partial_{x'}\phi',\]
up to a relative error $O(\eps v)$. Making the same argument in the other direction, suppose we begin with a coupling constant $b'$, and accelerate a particle from rest to a velocity $\vec{u} = w^1\hat{e}_1$. Then the dynamics from Sections~\ref{sec:oscillation} and \ref{sec:radiation} hold. The field radiates continuously from the particle's point of origin with wavenumbers $\lesssim k_{dB}$, and the particle oscillates at the frequency $\gamma^{-1}\omega_c$. However, the same transformation shows that, in our original reference frame, the particle started with a velocity $\vec{u}$ and ended with a velocity $\vec{u}+\vec{w}$, and the Lorentz covariance of the wave four-vector recovers the appropriate wavelength and \linebreak  oscillation frequency.

We recover the steady-state wavepacket (\ref{eq:steadystategeneral}) for free in this derivation. Indeed, if the particle undergoes \emph{no} Zitterbewegung, the transformed fixed-velocity state $\vec{u}$ satisfies the new trajectory equation exactly. Since the field equation is exactly Lorentz covariant, the transformed wavepacket remains a solution in the new reference frame.

Before moving onto transverse Lorentz transformations, we note that a core weakness of this argument is that we \emph{cannot} boost into the particle's post-acceleration frame of reference. Our dynamical harmony of phases does \emph{not} extend to the case in which the particle is spontaneously brought to rest. Study of this case is outside the scope of the current work.

Now, transverse Lorentz boosts leave the derivative $\partial_x\phi = \partial_{x'}$ unchanged, so our inline dynamics continue to satisfy the equation
\[d_{t'}\left(m\gamma' u'\right) = (\gamma')^{-1} mb(\partial_{x'}\phi(q_t)).\]
The only (indirect) effect is that the $\partial_{t'}\phi$ term now introduces a particle oscillation of the order $O(\eps v)$ in the transverse direction. In turn, the particle now experiences $\nabla\phi$ from slightly off the axis of symmetry, reducing the derivative $\partial_{x'}\phi(q_t)$ by a relative error $O(\eps^2v^2)$. Applying a similar argument as before shows that, if we start a particle at rest in the transformed frame and accelerate it in the $\hat{e}_1$ direction, the particle continues to vibrate in the direction $\hat{e}_1$ at the frequency $\gamma^{-1}\omega_c$, and is now forced in the transverse direction at the same frequency. The continuous source of radiation at the particle's point of origin is as before. Unlike the inline direction, however, there is no rescaling of either $b$ or $\phi$.

The difference of scaling between transverse and in-line Lorentz boosts means that, under a general Lorentz transformation, $b$ becomes anisotropic. If we break a generic Lorentz boost into in-line and transverse directions, the rescaled value $\tilde{b}$ governs dynamics in the in-line direction and the value $b$ governs dynamics in the orthogonal direction. This anisotropy does not affect the form of the wavepacket (\ref{eq:steadystategeneral}) or the wavelength and frequency seen in the dynamical harmony of phases. Nevertheless, it \emph{does} affect the amplitude of the particle's vibration in each direction. A more detailed study of this vibration amplitude is outside the scope of the present study.

\begin{adjustwidth}{-\extralength}{0cm}

\reftitle{References}

\PublishersNote{}
\end{adjustwidth}
\end{document}